\documentclass[a4paper,openany,12pt]{book}
\usepackage{mathptmx}
\usepackage{graphicx}
\usepackage{chapterbib} %
\usepackage{color}
\newcommand{\chapterauthor}[1]{\textsc{#1}\section*{}}  %

\title{Guided wave optics as a testbed for extreme waves}
\author{}
\begin{document}
\frontmatter
\maketitle
\tableofcontents
\mainmatter
\chapter[Extreme events in forced oscillatory media...]{Extreme events in forced oscillatory media in 0, 1 and 2 dimensions}

\chapterauthor{S.~Barland, M.~Brambilla, L.~Columbo, B.~Garbin, C.J.~Gibson, M.~Giudici, F.~Gustave, C.~Masoller, G.-L.~Oppo, F.~Prati, C.~Rimoldi, J.R.~Rios, J.R.~Tredicce, G.~Tissoni, P.~Walczak, A.M.~Yao, J.~Zamora-Munt}

\section{Abstract}

One of the open questions in the field of optical rogue waves is the relevance of the number of spatial dimensions in which waves propagate. Here we review recent results on extreme events obtained in 0, 1 and 2 spatial dimensions in the specific context of forced oscillatory media. We show that some dynamical scenarii can be relevant from 0 to 2D while others can take place only in sufficiently large number of spatial dimensions.

\section{Introduction}
A decade after the publication of the seminal paper on optical rogue waves \cite{solli2007optical}, a whole new field of study has emerged, focussing on extreme waves in optical systems. While this first study relied obviously heavily on the use of the nonlinear Schrödinger equation (NLSE) to describe the envelope of water waves and that of optical waves, the field has now considerably broadened (see eg \cite{akhmediev2016roadmap} for a recent overview). At this point, optical systems have become an excellent playground for the analysis not only of the optical equivalent of oceanic rogue waves as studied in the NLSE but more generally for the study of large and statistically (thus in some sense loosely) defined events. Thanks to this broader vision, the relevance of the field has increased a lot, in particular for the emphasis put on individuating the general physical or dynamical concepts which can lead to the emergence of heavy tailed statistics. Examples of such mechanisms (many of which are reviewed in this volume) include solitons in integrable systems \cite{akhmediev2009rogue,kibler2010peregrine,kedziora2013classifying,suret2016single} and their dissipative counterparts \cite{lecaplain2012dissipative,rimoldi2017spatiotemporal},  spatiotemporal chaos \cite{Selmi16,coulibaly2017extreme}, joint role of disorder and nonlinearity \cite{pierangeli2015spatial,pierangeli2016turbulent} or modulational instability \cite{mussot2009observation,dudley2009modulation}. Of course one key question since the very early days of the field is the role of spatial dimensionality of the system under study in the emergence of extreme waves. This issue is presently explicitely addressed in the context of optical fibers (see chapter by Tonello \textit{et al} in this volume) and photorefractive media (see chapter by Pierangeli \textit{et al} in this volume). Here we address this issue by reviewing conceptually similar systems consisting of forced oscillatory media (namely a laser with optical external forcing or an optical parametric oscillator) with different number of spatial dimensions. In the first section, we review experimental and numerical observations performed on a physical device whose spatial dimensions are very small, strongly constraining the spatial degrees of freedom of the slowly varying envelope of the field to the point that the experiment can be modelled with a set of ordinary differential equations \cite{Bonatto_PRL_2011,Zamora_PRA_2013}. In this case, the only source of complexity (besides possible stochastic effects) is that of low dimensional chaos. In the second section, the experimental arrangement is such that the forced laser is emitting on many longitudinal modes, \textit{i.e.} it is a 1-dimensional system, spatially extended along the propagation direction \cite{gustave2015dissipative,gustave2016phase,gustave2017formation}. In the third section, we present a theoretical analysis of a 2-dimensional device, spatially extended in the dimensions transverse to propagation, where spatial coupling occurs via diffraction \cite{Oppo13,Gibson16}.

\section{0D}\label{sec:0d}
\begin{figure}[h!]
\begin{center}
\centering\includegraphics[width=12cm]{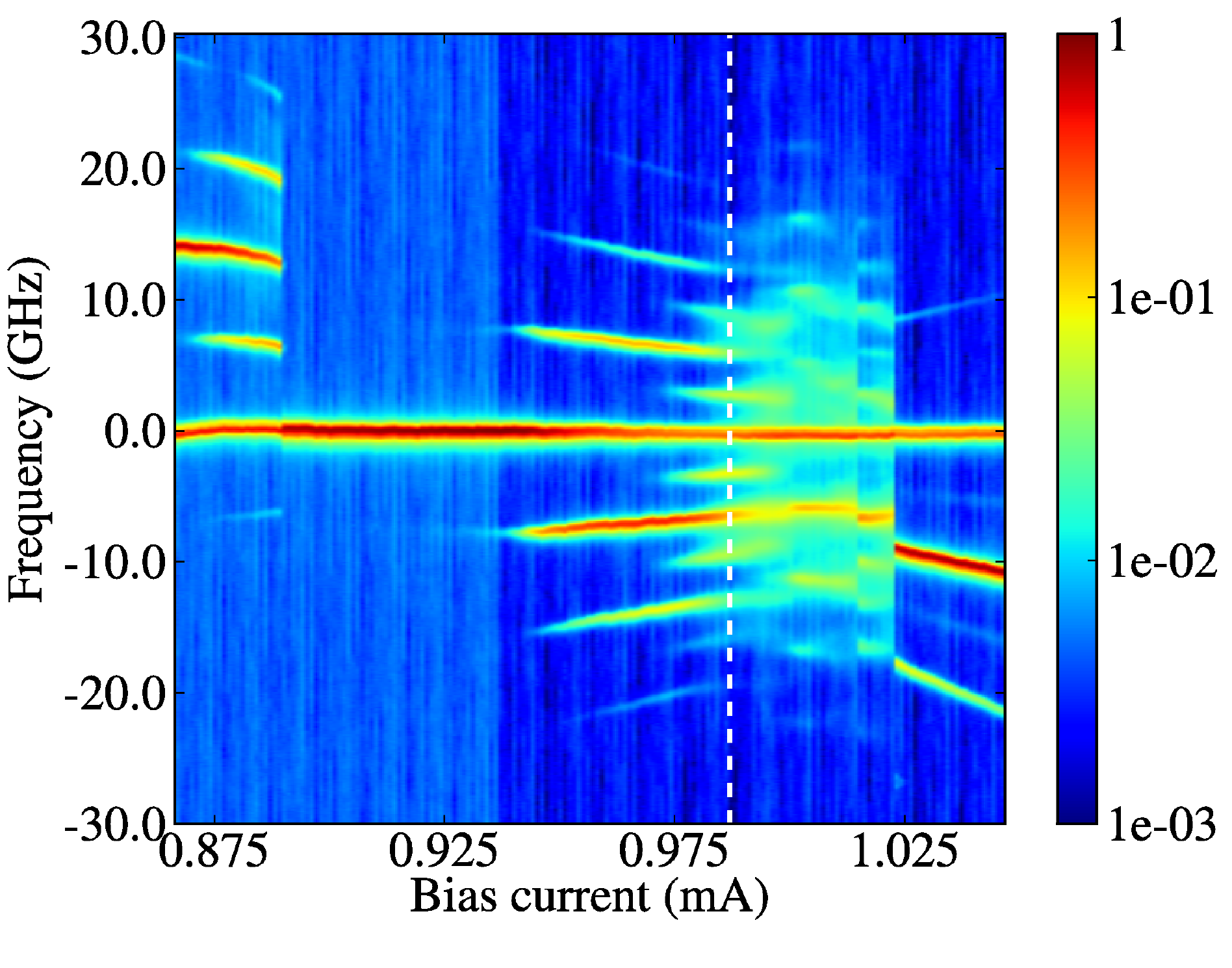}

\end{center}
\caption{Experimental bifurcation diagram displaying the optical
spectra of the slave laser vs. the bias current. The dashed line
denotes the zero detuning condition between the master and the slave
lasers.}
\label{0Dexpe_1}
\end{figure}

In this section we discuss the generation of rogue waves in a simple
system composed by a semiconductor laser with continuous-wave
optical injection.

A single-transverse mode Vertical-Cavity Surface-Emitting
semiconductor Laser (VCSEL) is injected by a wavelength-tunable
beam. VCSEL operates at around 980 nm and it features threshold
current at about 0.2 mA. The laser is pumped by a current supply
stabilized to better than 0.1 $\mu$A and its package is temperature
stabilized to better than $0.01~$K. It is important noting that the
VCSEL emits a single polarization mode in the parameters range
explored in our experiment.  The optical injection beam can be tuned
in the range 960-980 nm by steps of 51 GHz and, within each step, it
can be tuned continuously on a range of 6~GHz. Its power can be
adjusted up to 2 mW. The slave laser output passes through an
optical isolator preventing from back reflections from the detection
part of the setup. Intensity output is monitored by a detection
apparatus consisting of a 6 GHz oscilloscope (LeCroy Wavemaster) and
an amplified photodetector covering the 0.01-4.2~GHz frequency range. The
optical spectra are detected with a scanning Fabry-Perot
interferometer (free spectral range of 78 GHz and Finesse of 125).

Critical parameters of our experiment are the frequency detuning
between the injected beam ($\nu_o$) and the slave lasers ($\nu_s$),
$\Delta \nu=\nu_o-\nu_s$, the injection intensity, $P_{inj}$, and
the slave laser pumping current, $J$. It is worthwhile noting that
these parameters are not independent. A variation of $J$, for
example, affects $\nu_s$ via the change of semiconductor refraction
index, thus inducing a red shift of the slave laser emission for
increasing $J$ which is, for the laser used, 116 MHz/$\mu$A for current values around
$J$=1~mA. In Fig. \ref{0Dexpe_1} we show the optical spectrum
emission of the slave laser as a function of $J$ varied from
0.867~mA to 1.046~mA. This current variation implies a variation of
$\Delta\nu$ between -14 GHz and +7 GHz, with $\Delta\nu=0$ occurring
at $J$=0.987~mA (indicated with a dashed line on the graph).

We obtain extreme events in the intensity output of the slave laser
in the parameter region where the slave laser gets unlocked from the
master laser and develops a chaotic state through a period doubling
bifurcation sequence ($J \approx 0.975$ mA).

\begin{figure}[tbp]
\centering\includegraphics[width=14cm]{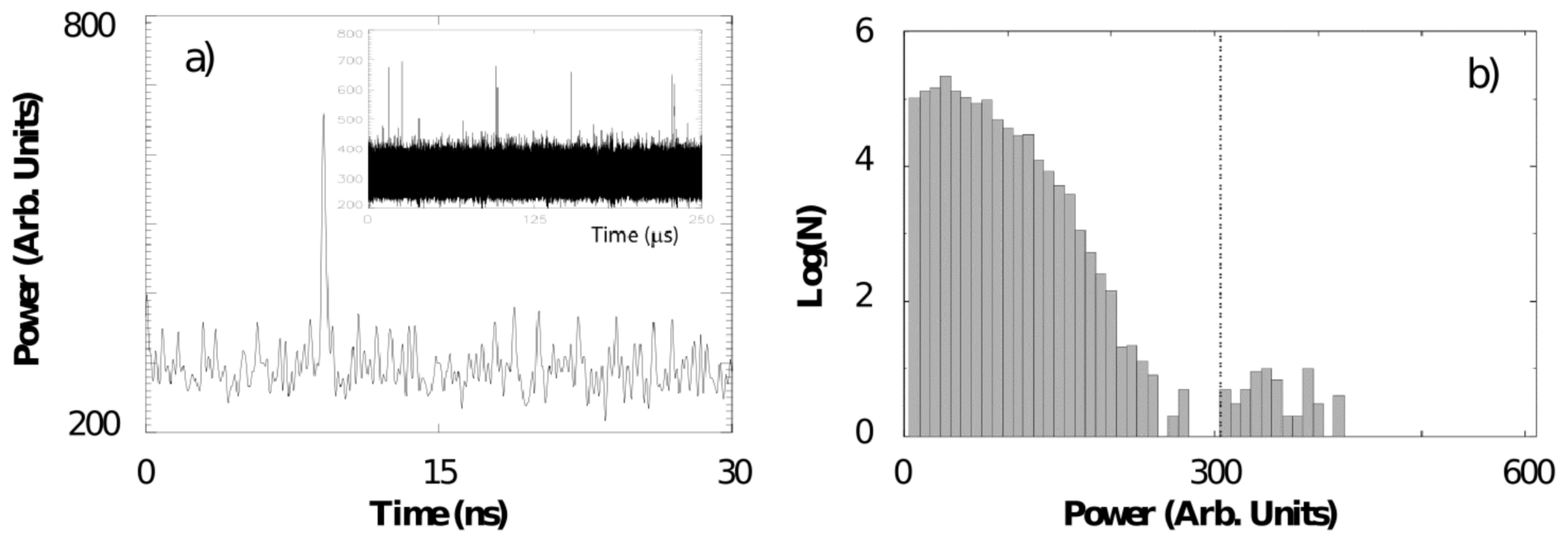} \caption{(a)
Typical time series of the slave laser intensity, where rare large
pulses can be observed. The experimental parameters are $J$=0.976
mA, $\Delta\nu$=-1.3 GHz and $P_{inj}=1.1$ mW. (b) Power
distribution of the number $N$ of pulses emitted by the slave laser.
The vertical dashed line denotes the mean value plus eight standard
deviations and thus mark the limit for a pulse to be considered a
rogue event. The histogram in (b) corresponds to the time series
displayed in (a)} \label{0Dexpe_2}
\end{figure}

In this region the intensity output of the slave laser shows a
series of small pulses with sporadic extremely high ones, as shown
in Figure \ref{0Dexpe_2}(a).  The corresponding power distribution
of these events exhibits a long tail, thus revealing the presence of
extreme events. In Figure \ref{0Dexpe_2}(b) we show one typical
pulse histogram obtained where a number of events crosses the limit
above which a pulse can be considered extreme. By increasing the
slave current value the system is pushed deeper in the chaotic
region and the probability of large power pulses emission increases.
Accordingly, the mean value of the peak power increases and the relative weight of the distribution decreases, to the point that no pulses overcome the "rogue waves" threshold anymore.

This bifurcation scenario and, more specifically, the presence of
these extreme events is found in a wide range of optically injected
power (from 0.6 to 1.4 mW). The critical detuning for which these
events occur tends to zero as $P_{inj}$ decreases and it can also be
positive for low injected power. The range of injected power where
RWs are observed depends also on the slave laser pumping current. If
$J$ is increased to 1.321 mA this range is shifted upwards (from 1.0
to 2.0 mW) while the critical detuning remains almost unchanged.
Finally, for $J>2$~mA the chaotic region supporting extreme events disappears.

These experimental results are well described by a rate-equations
model for the slow envelope of the complex electric field $E$ and
for the carrier density $N$ \cite{phys_rep_2005}.

\begin{eqnarray}
  \frac{dE}{dt} &=& \kappa(1+i\alpha)(N-1)E +i\Delta\omega E + \sqrt{P_{inj}}+ \sqrt{D}\xi(t), \nonumber \\
  \frac{dN}{dt} &=& \gamma_N(\mu(t)-N-|E|^2),
    \label{eq:model_MS_RWs}
\end{eqnarray}
where $\kappa$ is the field decay rate, $\alpha$ is the line-width enhancement factor, $\gamma_N$ is the carrier decay rate and $\mu$ is the injection current parameter, normalized such that the threshold current of the free-running laser is $\mu_{th}$=1. $\Delta\nu=\Delta \omega/2\pi$ is the optical frequency detuning, $\Delta \omega=\omega_0-\omega_s$, with $\omega_0$ and $\omega_s$ being the frequencies of the master and slave lasers respectively. $P_{inj}$ is the constant optical injection strength. The term $D\xi(t)$ represents optical noise: $\xi$ is a complex uncorrelated Gaussian white noise and $D$ is the noise strength.

The model equations were numerically solved using the same parameters as in \cite{Bonatto_PRL_2011,Zamora_PRA_2013}: $\kappa=300$ n$s^{-1}$, $\alpha=3$, $\gamma_N=1$ ns$^{-1}$, $P_{inj}=60$ ns$^{-2}$, and the other parameters were varied.

\begin{figure}[t]
\centering
\includegraphics[width = 0.49\columnwidth]{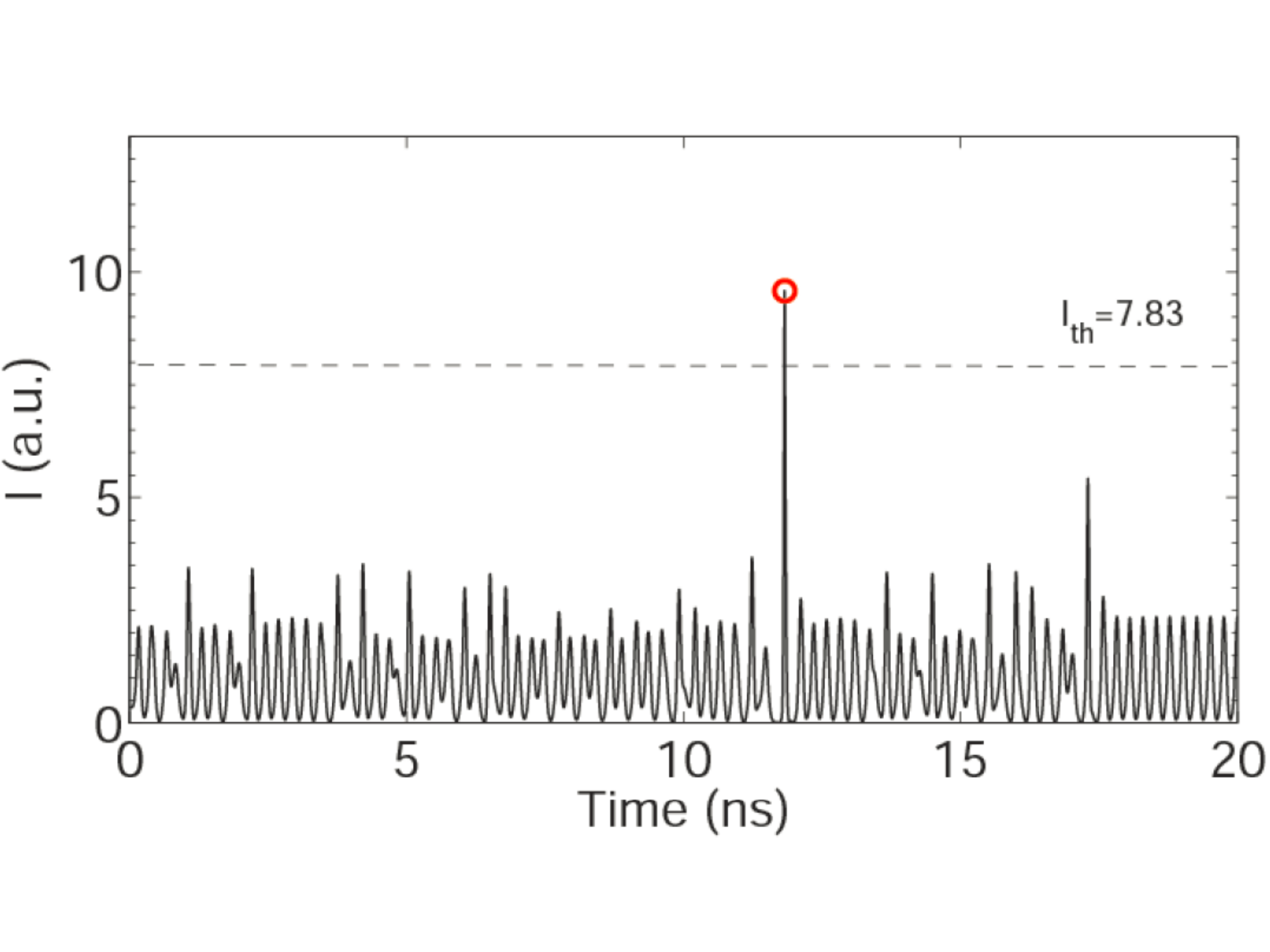}
\includegraphics[width = 0.49\columnwidth]{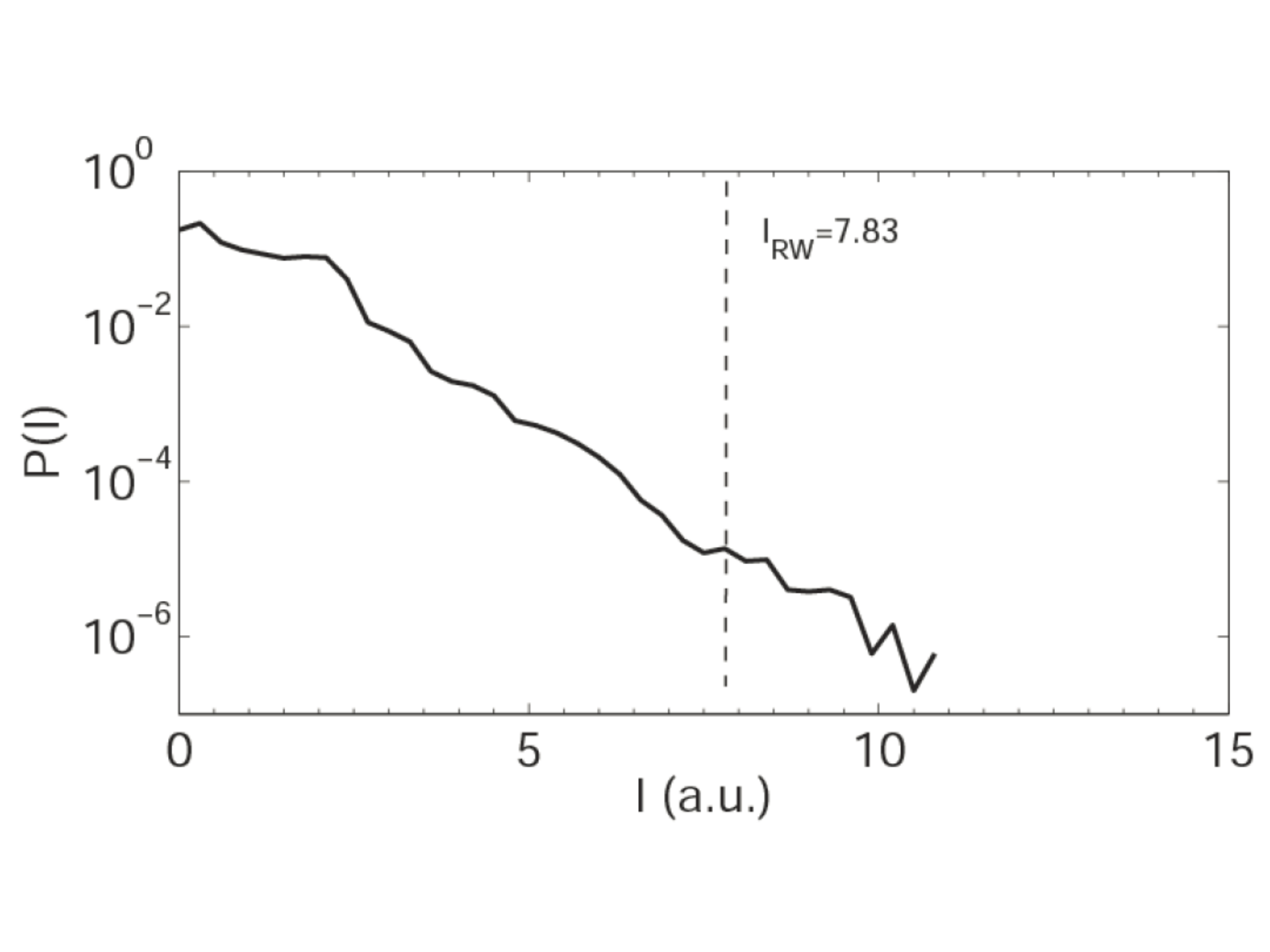}
\caption{Deterministic rogue wave. (Left) Time trace of the laser intensity. The cycle indicates the detection of a RW. (Right) Probability distribution function (pdf) of intensity values. The vertical dashed line indicates the RW threshold, $\left\langle I \right\rangle + 8 \sigma_I$. Parameters are as in \cite{Bonatto_PRL_2011}, point B: $\kappa=300$ n$s^{-1}$, $\alpha=3$, $\gamma_N=1$ ns$^{-1}$, $P_{inj}=60$ ns$^{-2}$, $\Delta \nu=-1.86$ GHz, $\mu=1.96$ and $D=0$.
}
\label{fig:time_trace_pdf}
\end{figure}

Figure~\ref{fig:time_trace_pdf} displays the time trace of the laser intensity, $I=|E|^2$, generated from random initial conditions, and the corresponding probability distribution function (PDF) of intensity values. We define a RW event when the intensity is higher than a threshold given by $\left\langle I \right\rangle + 8 \sigma_I$, where $\left\langle I \right\rangle$ and $\sigma_I$ are the mean and the standard deviation of the intensity pdf.

\begin{figure}[htb]
\centering
\includegraphics[width = 0.32\columnwidth]{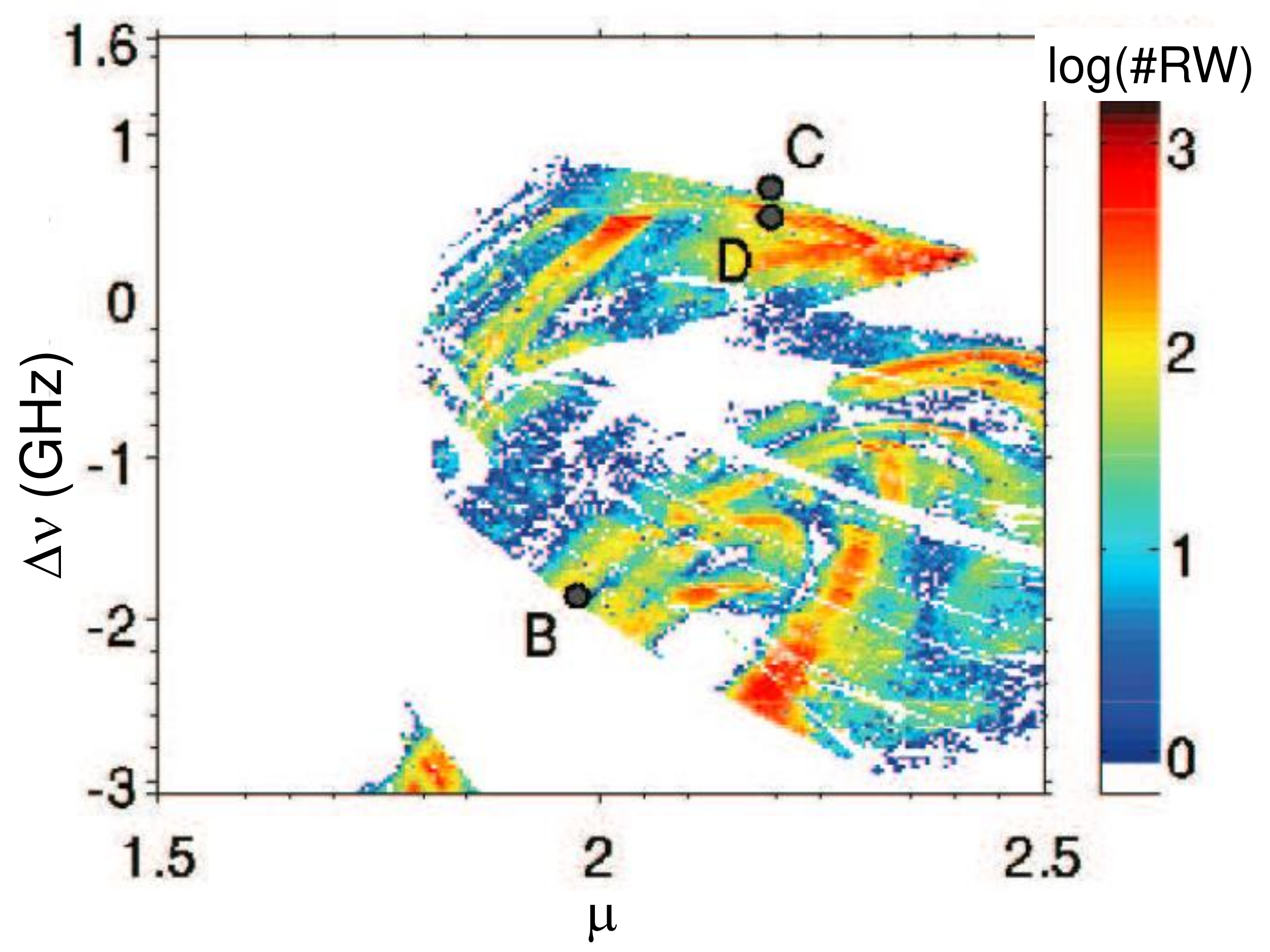}
\includegraphics[width = 0.32\columnwidth]{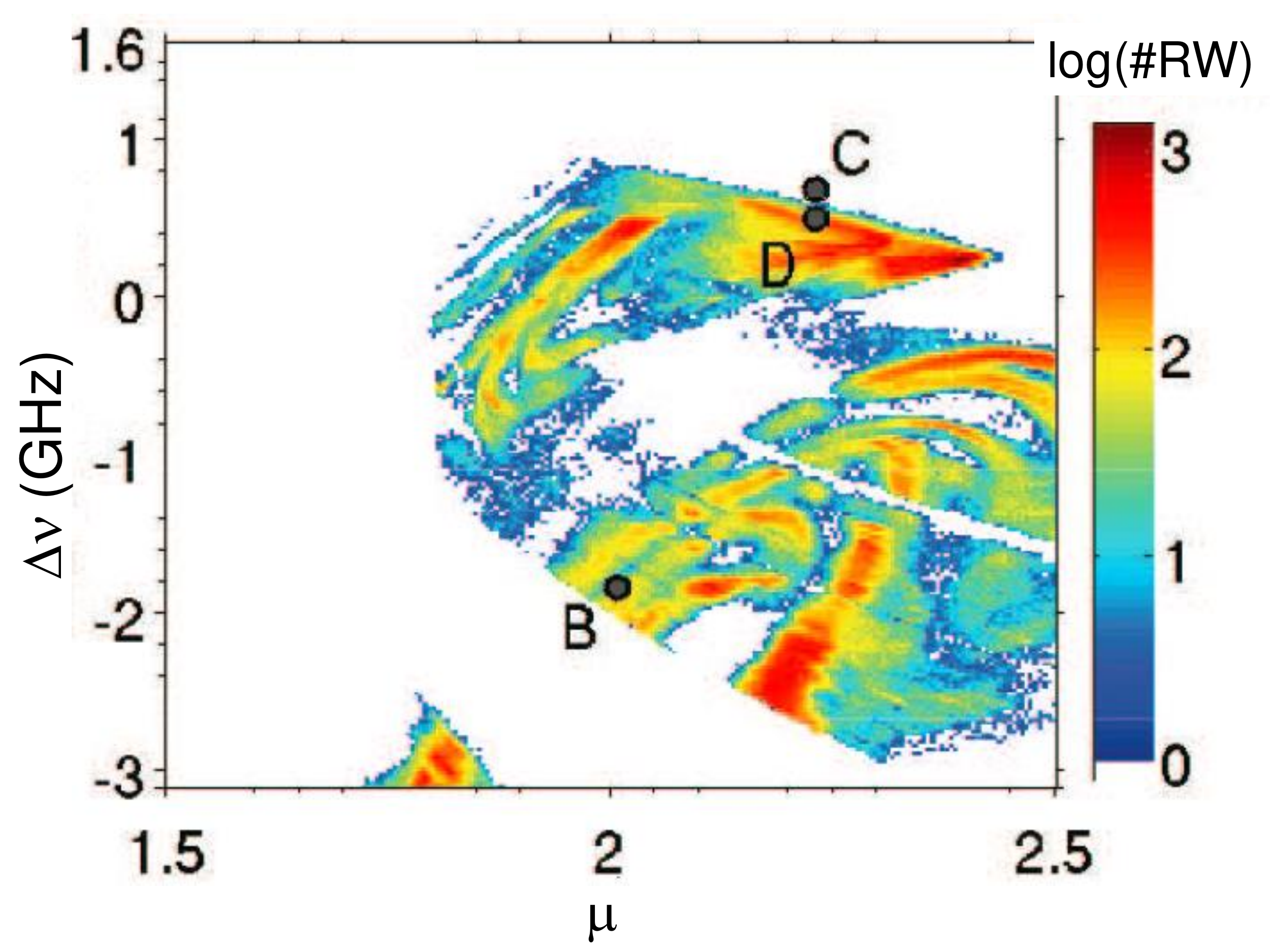}
\includegraphics[width = 0.32\columnwidth]{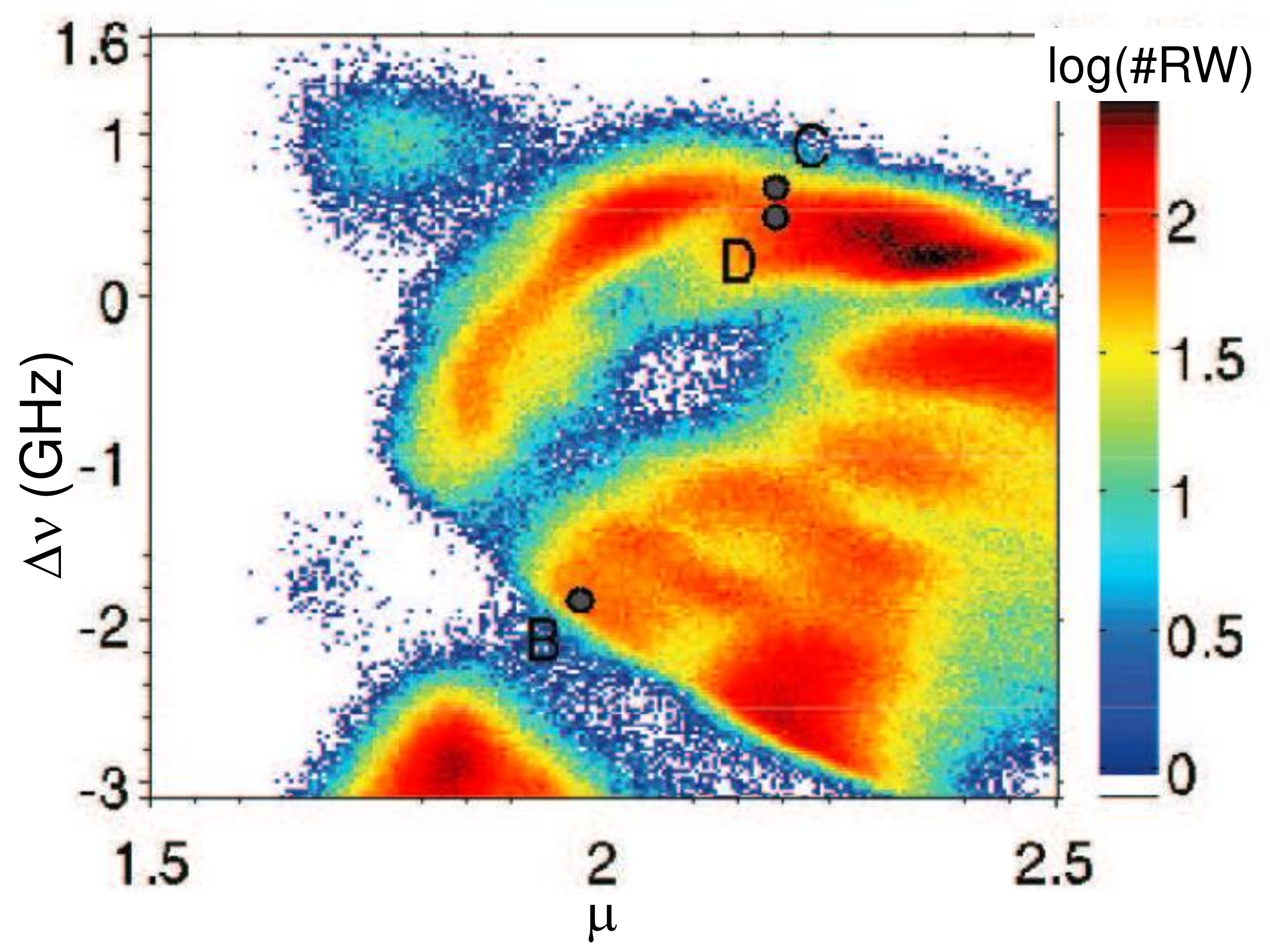}
\caption{Number of RWs in the parameter space $(\mu,\Delta \nu)$ for different noise strengths. The color code is plotted in logarithmic scale in order to increase the contrast of the regions with a small number of RWs. In the white regions no RWs are detected. Left: $D=0$, center: $D=10^{-4}$ ns$^{-1}$, right: $D=10^{-2}$ ns$^{-1}$. The set of parameters used in Figs.~\ref{fig:time_trace_pdf},~\ref{fig:predicta} are labeled as B (as in \cite{Bonatto_PRL_2011}: $\mu=1.96$, $\Delta \nu=-1.86$ GHz); labels C and D indicate a boundary region: in point D there are deterministic RWs, while in C, RWs are induced by strong  noise.
}
\label{fig:RW_number}
\end{figure}
\begin{figure}[htb]
\centering
\includegraphics[width = 0.49\columnwidth]{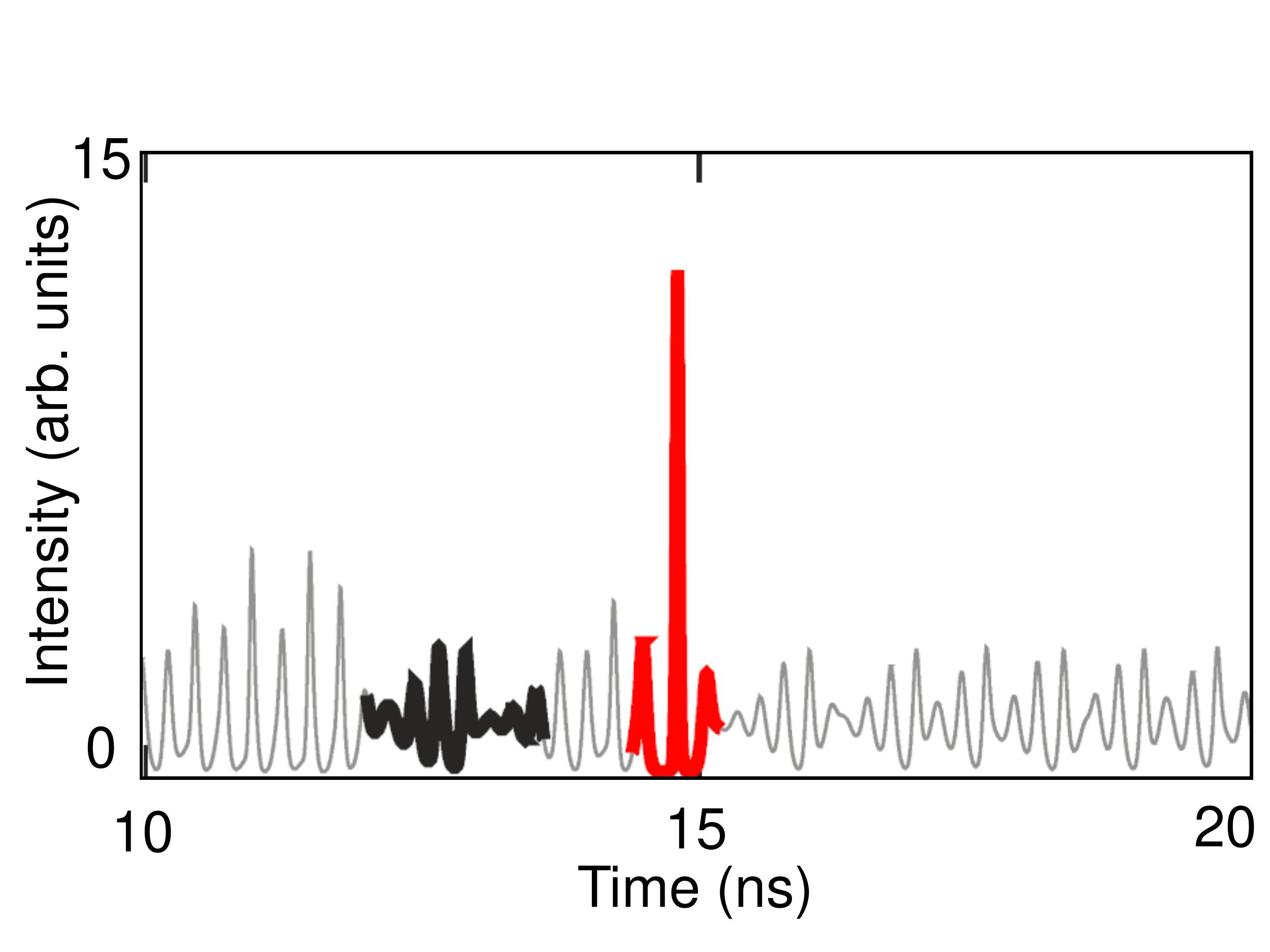}
\includegraphics[width = 0.49\columnwidth]{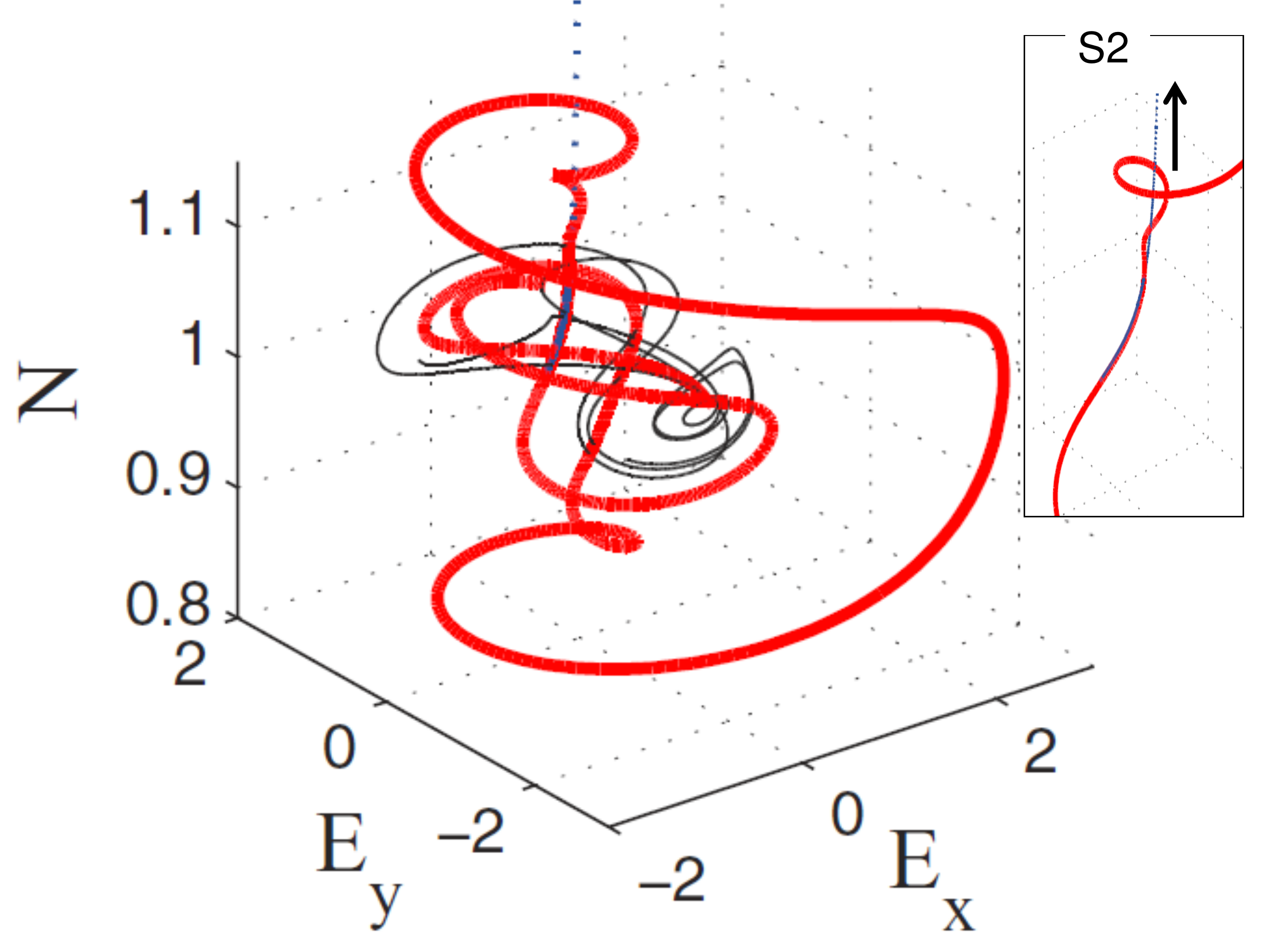}
\caption{Left: intensity time trace displaying small oscillations in the dense part of the chaotic attractor (in black) and in the sparse part of the attractor, when a RW occurs. Right: plot of the two trajectory segments in the $(E_x,E_y, N)$ phase space. The inset displays in detail the evolution along the stable manifold of S2 (dotted blue).
}
\label{fig:phase_space}
\end{figure}
\begin{figure}[htb]
\centering
\includegraphics[width = 0.49\columnwidth]{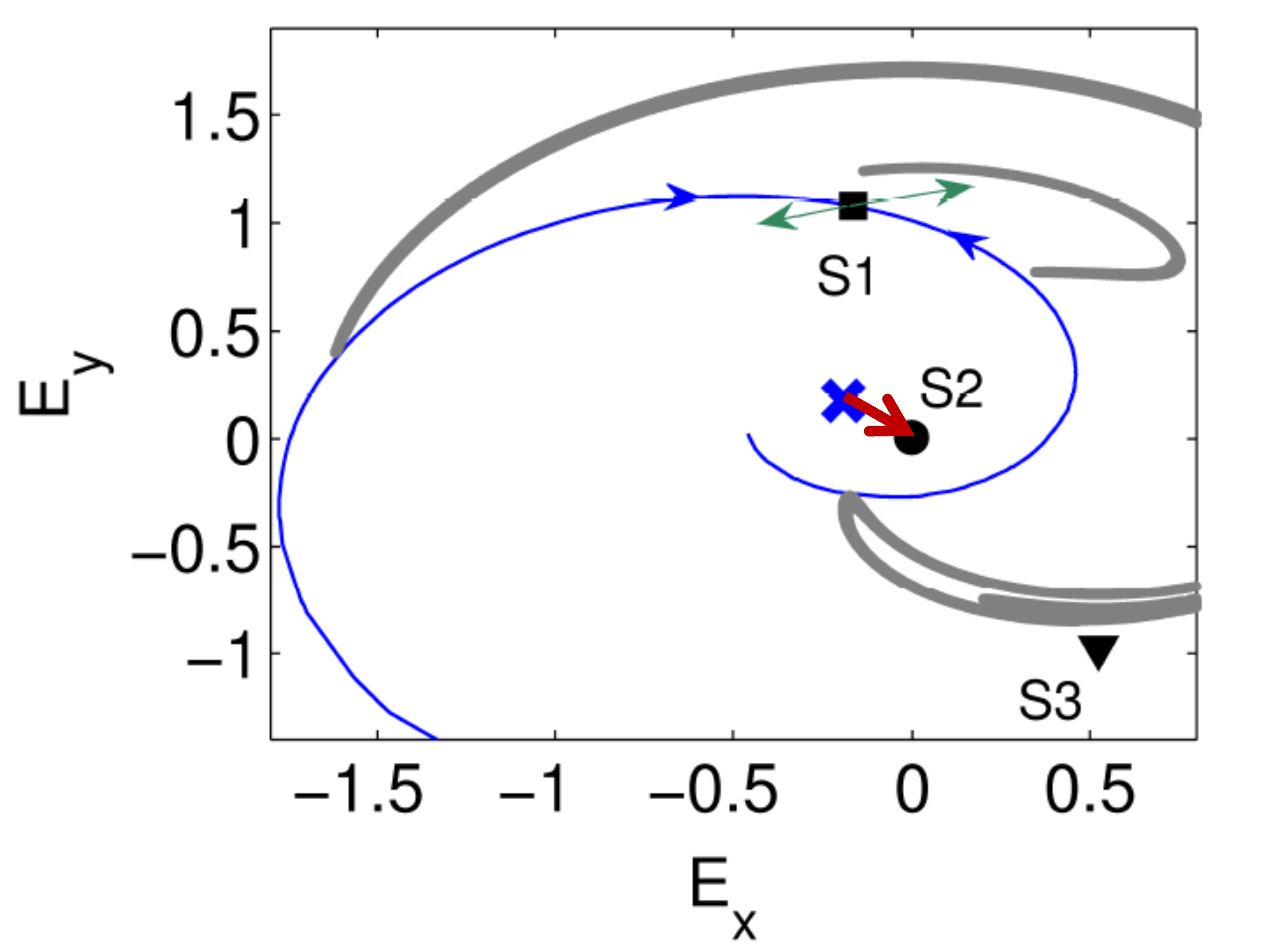}
\includegraphics[width = 0.49\columnwidth]{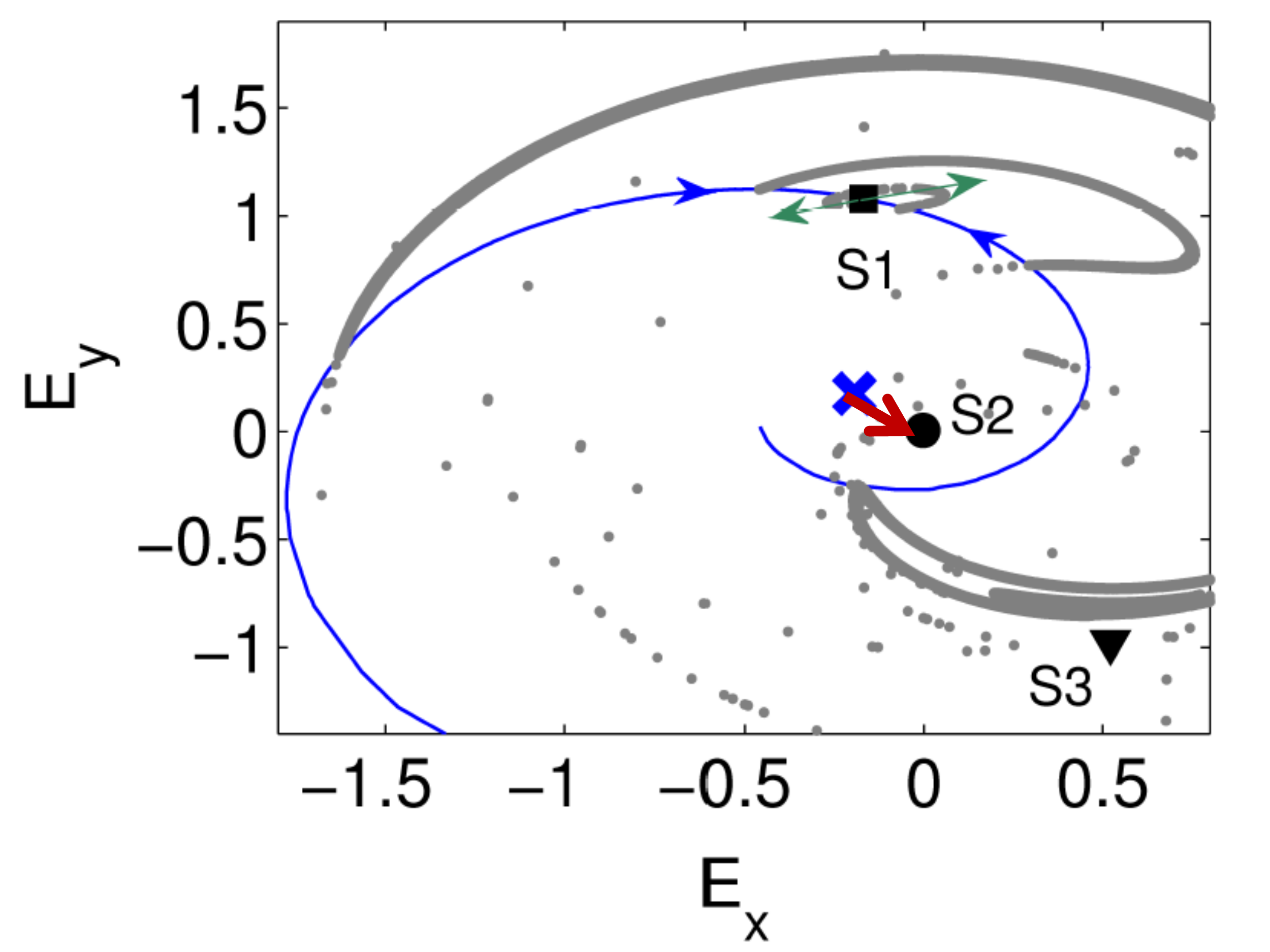}
\caption{Poincar\`{e} section of the attractor with the plane $N = 1.0036$. The three fixed points (S1, S2, S3) and the stable manifold of S1 are also projected in the Poincar\`{e} plane. Left: before the collision with the stable manifold of S1, the attractor (developed from S3) is confined to a region of the phase space that does not include the stable manifold of S2. The parameters are $\mu=2.2$, $\Delta \nu=0.6$ GHz. Right: after the collision the attractor expands and gains access to the region where the stable manifold of S2 is located. The parameters are $\mu=2.2$, $\Delta \nu=0.594$ GHz.
}
\label{fig:poincare}
\end{figure}
\begin{figure}[htb]
\centering
\includegraphics[width = 0.32\columnwidth]{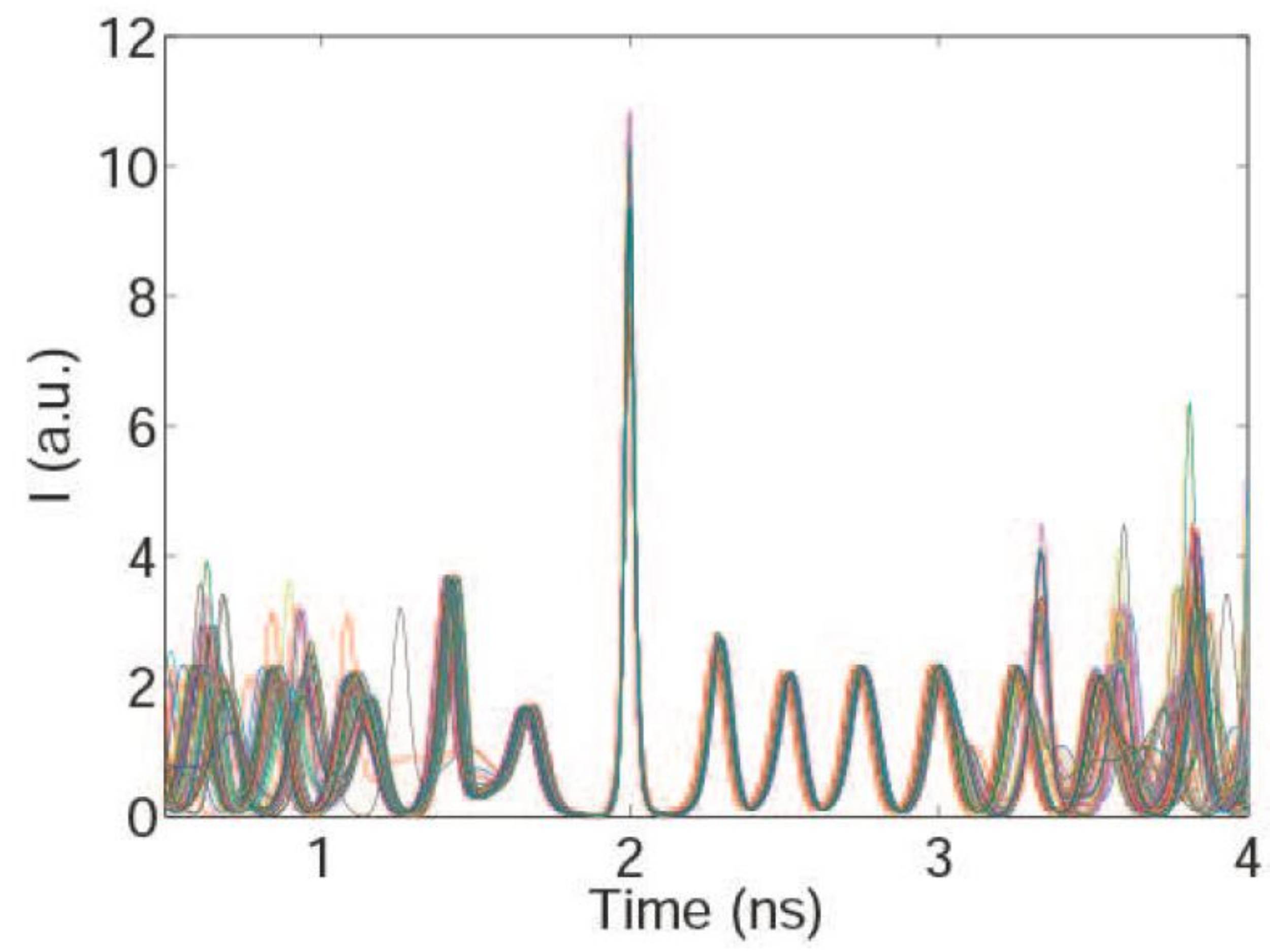}
\includegraphics[width = 0.32\columnwidth]{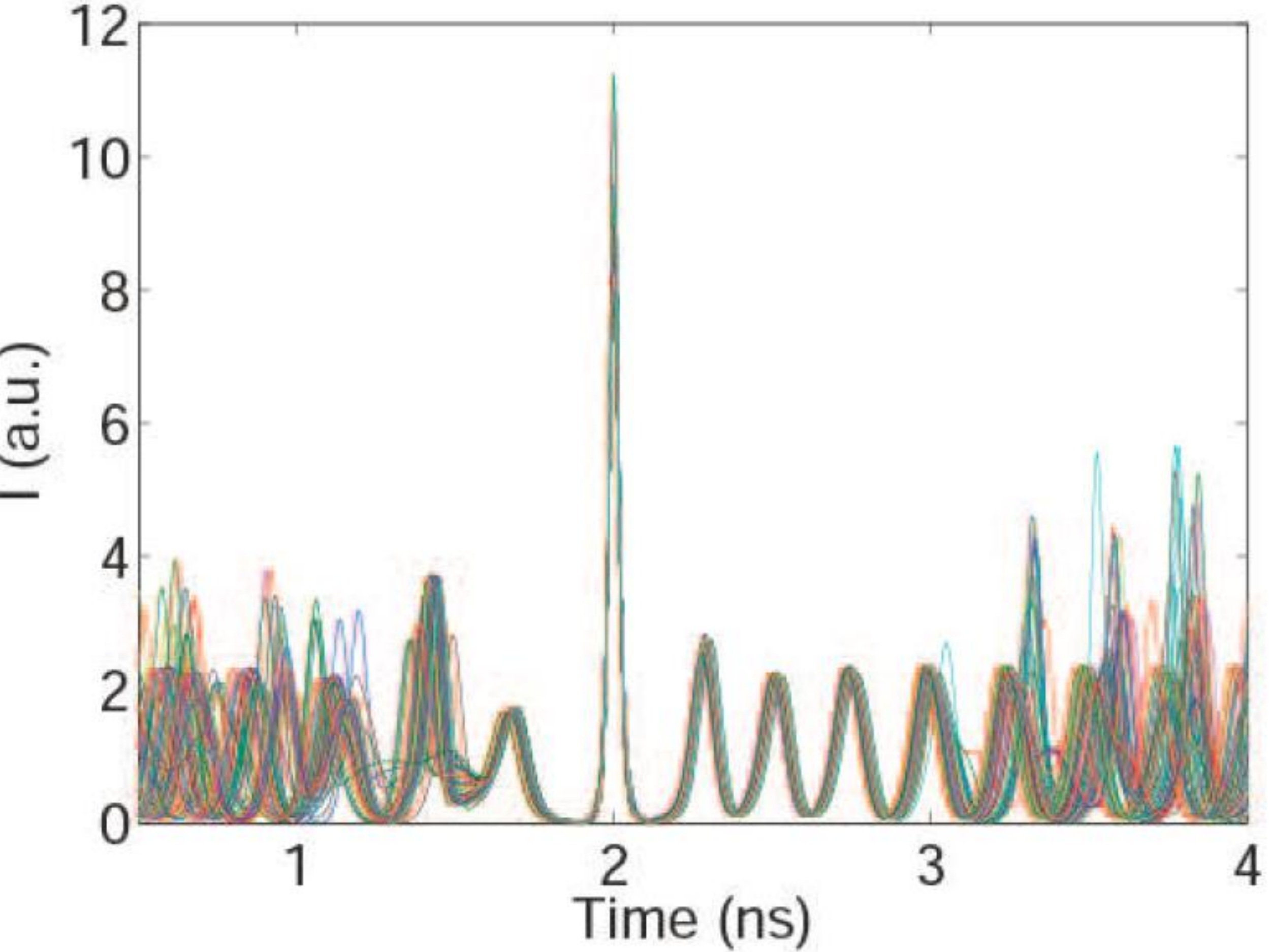}
\includegraphics[width = 0.32\columnwidth]{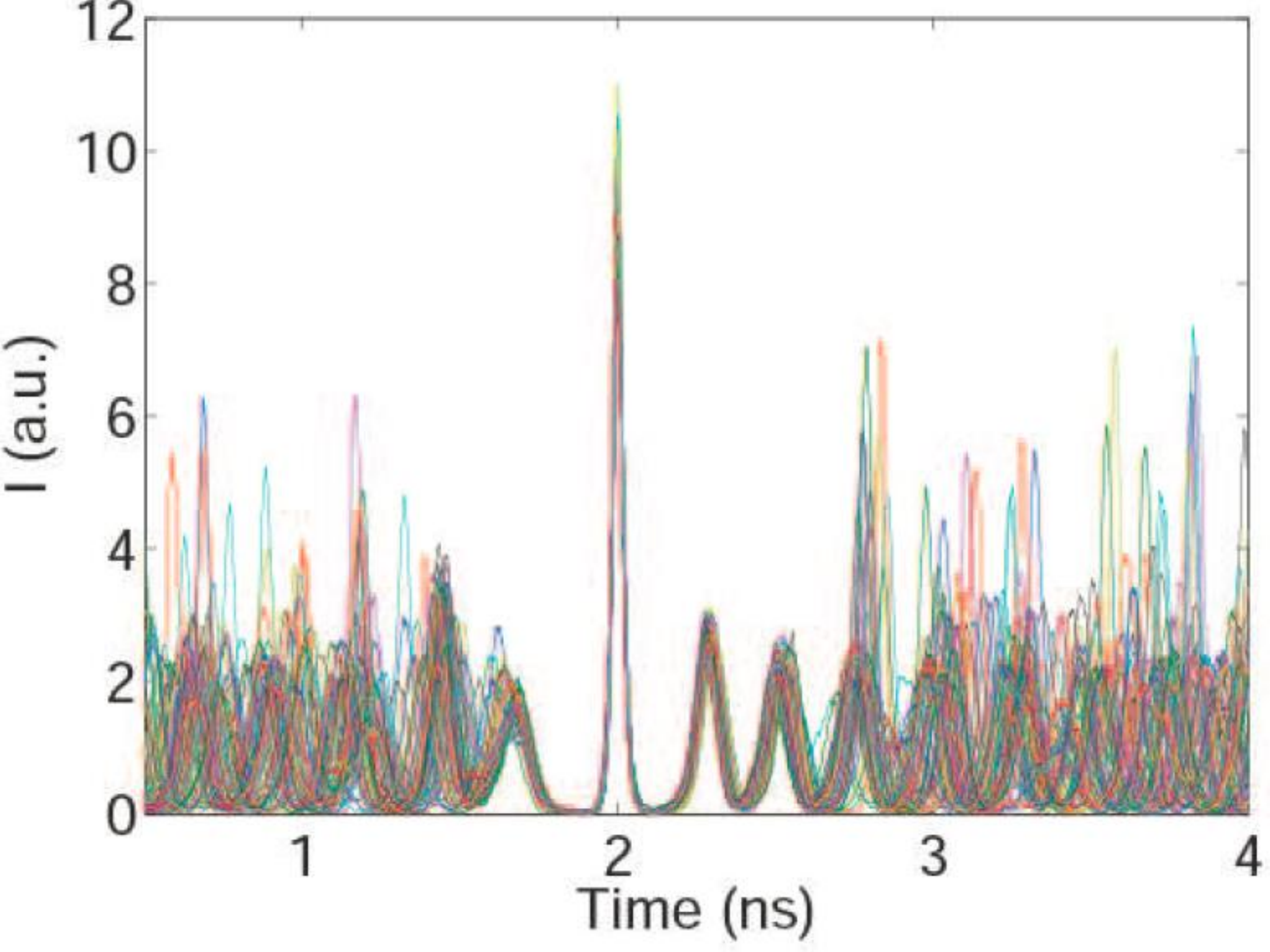}
\caption{Time traces of the intensity centered at the maximum of the RW for different noise strengths. Left: $D=0$, center: $D=10^{-4}$ ns$^{-1}$, right: $D=10^{-2}$ ns$^{-1}$. The parameters are $\mu=1.96$, $\Delta \nu=-1.86$ GHz (point B).
}
\label{fig:predicta}
\end{figure}

As shown in \cite{Bonatto_PRL_2011}, RWs appear in regions of the parameter space where the dynamics is chaotic. However, that is not enough to observe RWs as there are regions of chaotic dynamics with RWs and others without RWs.

In our model, in the deterministic case ($D=0$), RWs appear abruptly when a parameter is varied, as it can be seen in Fig.~\ref{fig:RW_number}(a). Then, two relevant questions are: which special properties the chaotic regions with RWs have? and what type of bifurcations produce RWs?

As shown in \cite{Zamora_PRA_2013}, RWs can not occur at any point of the phase space. Only a small region of the phase space can lead to a RW when it is visited by the trajectory. In the chaotic regions with RWs, the trajectory can access a narrow region of the phase space (referred to as ``the rogue wave door'') where the stable one-dimensional manifold of a saddle point (referred to as S2) is located. Whenever the trajectory closely approaches this region, a RW is likely to occur. The mechanism is as follows: the trajectory evolves along the stable manifold towards S2, which is the low-intensity solution with high $N$ and $I=|E^2|\sim 0$. However, S2 has a two-dimensional unstable manifold, and thus, the trajectory also spirals out. Because during the evolution along the stable manifold of S2 the system accumulates carriers while approaching a low intensity, a high pulse is emitted. This process is illustrated in Fig.~\ref{fig:phase_space} that displays the intensity time-trace and the corresponding plot in the $(E_x,E_y, N)$ phase space, during the frequently visited (``dense'') part of the attractor (in black) and also during the RW (when the trajectory performs a large excursion in phase space, in red). The inset shows that the trajectory moves along the stable manifold of S2 (dotted blue) while spiraling out. The duration of the evolution along the stable manifold of S2 is irregular because it varies with the initial ``entry'' conditions and is affected by numerical noise, and in consequence, there is a wide distribution of pulse heights, as shown in Fig.~\ref{fig:time_trace_pdf}(b).

In contrast, in the chaotic regions without RWs, the trajectory in phase space never approaches this dangerous region of the phase space where the RW door (the one-dimensional stable manifold of S2) is located.

The type of bifurcation was also analyzed in \cite{Zamora_PRA_2013} and numerical evidence suggested that RWs appear as the result of a collision of the attractor (that develops from a fixed point referred to as S3) with the stable manifold of another saddle-focus point (referred to as S1). As shown in Fig.~\ref{fig:poincare}, after this collision the attractor expands and gains access to the phase space region where the RW door is. In other words, the attractor generated from fixed point S3 collides with the stable manifold of S1, which produces an abrupt expansion of the attractor, which in turn enables access to the region of the phase space where the stable manifold of S2 is; then, when the trajectory approaches this manifold, a RW can be triggered.

In \cite{Zamora_PRA_2013} it was also shown that RWs can be predicted with a certain prediction horizon. This is shown in Fig.~\ref{fig:predicta}, where we plot a large number of time-traces, centered at the maximum intensity of the RW. Note that not only before the RW, but also after, the intensity behaves in a predictable way. Here we also show the robustness of the RWs to noise. Noise reduces the horizon of predictability, as it could be expected, but even with strong noise, there is a considerable interval of time when the RW can be predicted. Moreover, the height of the RW is mainly unaffected by the noise strength.

To summarize, in our system RWs can be deterministic but they can also be induced by noise in certain parameter regions. Both, deterministic and stochastic RWs can be predicted within a certain time horizon. We have also found that in the deterministic case, RWs occur when an abrupt expansion of the attractor enables access to a particular region of phase space. The three fixed points of the system are relevant for understanding the RW phenomenon: an attractor develops from one fixed point, the stable manifold of a second fixed point acts a barrier, while the stable manifold of the third fixed point acts as a ``RW door'': if the trajectory closely approaches this manifold, a RW can be triggered.

\section{1D}\label{sec:1d}In the following we analyze experimentally and numerically the emergence of extreme events in a laser with optical injection with one spatial dimension. This is achieved by performing an experiment conceptually similar to the one described in \ref{sec:0d} but, as a slave laser, we will use a strongly multi-longitudinal mode laser instead of using a strictly single longitudinal/transverse mode laser. As we shall see in the following, the role of 0D  attractors (i.e. the associated homogeneous stationary solutions) will still provide some resemblance to RW events observed in the previous section, although other phenomena will strongly differ, due to the spatial extension of the system

In 1D systems the spatiotemporal dynamics of the laser is appropriately described by the propagation equation of the coherent field although, heuristically,  the spatial profile of the field can be interpreted in terms of  nonlinear interaction and competition of several empty cavity modes, as ruled by salient features of the device, such as gain width and free spectral range.
The system, now infinite dimensional, may exhibit a wealth of dynamical behaviors including spatiotemporal chaos which is an ideal irregular substrate over which extreme events can be recorded as high intensity peaks.
In this instance we adopted a model for spatially (1D) extended lasers, based on a set of rate equations which, apart from the different notations, is identical to Eqs. (\ref{eq:model_MS_RWs}), with the addition in the equation for the electric field of a propagation trem containing the first derivative along the longitudinal coordinate $z$ and a spectral filter (diffusion) term containing the second derivative \cite{gustave2015dissipative,gustave2016phase}
\begin{eqnarray}
\frac{\partial E}{\partial z}+\frac{\partial E}{\partial t}-d\frac{\partial^2 E}{\partial z^2}&=&T\left[-(1+i\theta) E+y+(1-i\alpha)ED\right]\,,\label{eq:modelE}\\
\frac{\partial D}{\partial t}&=&b\left[\mu-D\left(1+|E|^2\right)\right]\,.\label{eq:modelD}
\end{eqnarray}
$E$ and $D$ are, respectively, the slowly varying envelope of the electric field and the excess of carrier density with respect to transparency. %
The cavity losses are represented by the mirror transmissivity $T$, $y$ is the injected field amplitude, $\theta$ is the frequency detuning between the injected field and the closest cavity mode, $b$ is the ratio of the roundtrip time to the carrier lifetime, $\alpha$ and $\mu$ have the same meaning as in Eqs. (\ref{eq:model_MS_RWs})

The diffusion term
introduces an effective damping of short--scale spatial modulations and its weight is determined by the $d$ coefficient which is proportional to the squared ratio of the cavity free spectral range to the gain linewidth.

\begin{figure}[tb]
\centering\includegraphics[width=12cm]{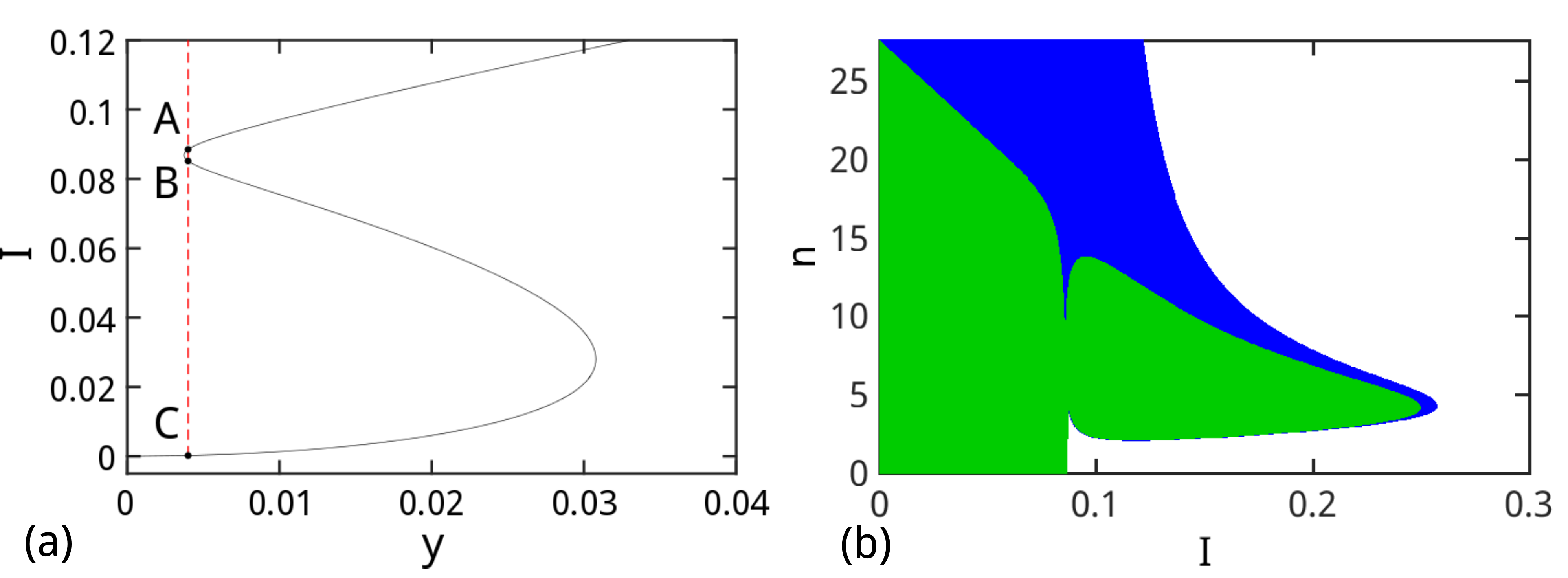}
\caption{(a) Stationary homogenous solution of Eqs. (\ref{eq:modelE}), (\ref{eq:modelD}) for $\mu=1.1$, $\theta=-3.04$, $\alpha=3$, $b=10$, $T=0.3$. Points $A$, $B$ and $C$ are the solutions corresponding to the value $y=0.004$ of the injected amplitude. (b) Instability domains in the $(I,n)$ plane according to Eqs. (\ref{eq:modelE}), (\ref{eq:modelD}) (in green) and for the pure rate equation model with $d=0$ (in blue).}
\label{fig:stat_curve}
\end{figure}

The stationary homogeneous solution $E_s=\rho_s\exp(i\phi_s)$, $D_s$ of Eqs. (\ref{eq:modelE}), (\ref{eq:modelD}) is
\begin{equation}\label{eq:stat}
y^2=\rho_s^2\left[\left(1-D_s\right)^2+\left(\theta+\alpha D_s\right)^2\right]\,,\quad
\phi_s=\arctan{\left(\frac{\theta+\alpha D_s}{D_s-1}\right)}\label{eq:phistat}\,,\quad
D_s=\frac{\mu}{1+\rho_s^2}
\end{equation}
and for parameters compatible with the experiment\footnote{Note that the pump threshold value for the pump parameter in the free running laser is $\mu=1$}, the stationary intensity curve $I=\rho_s^2$ of the electric field as a function of the injection amplitude $y$ is multivalued as shown in Fig. \ref{fig:stat_curve}(a).
\begin{figure}[tb]
\centering\includegraphics[width=0.8\textwidth]{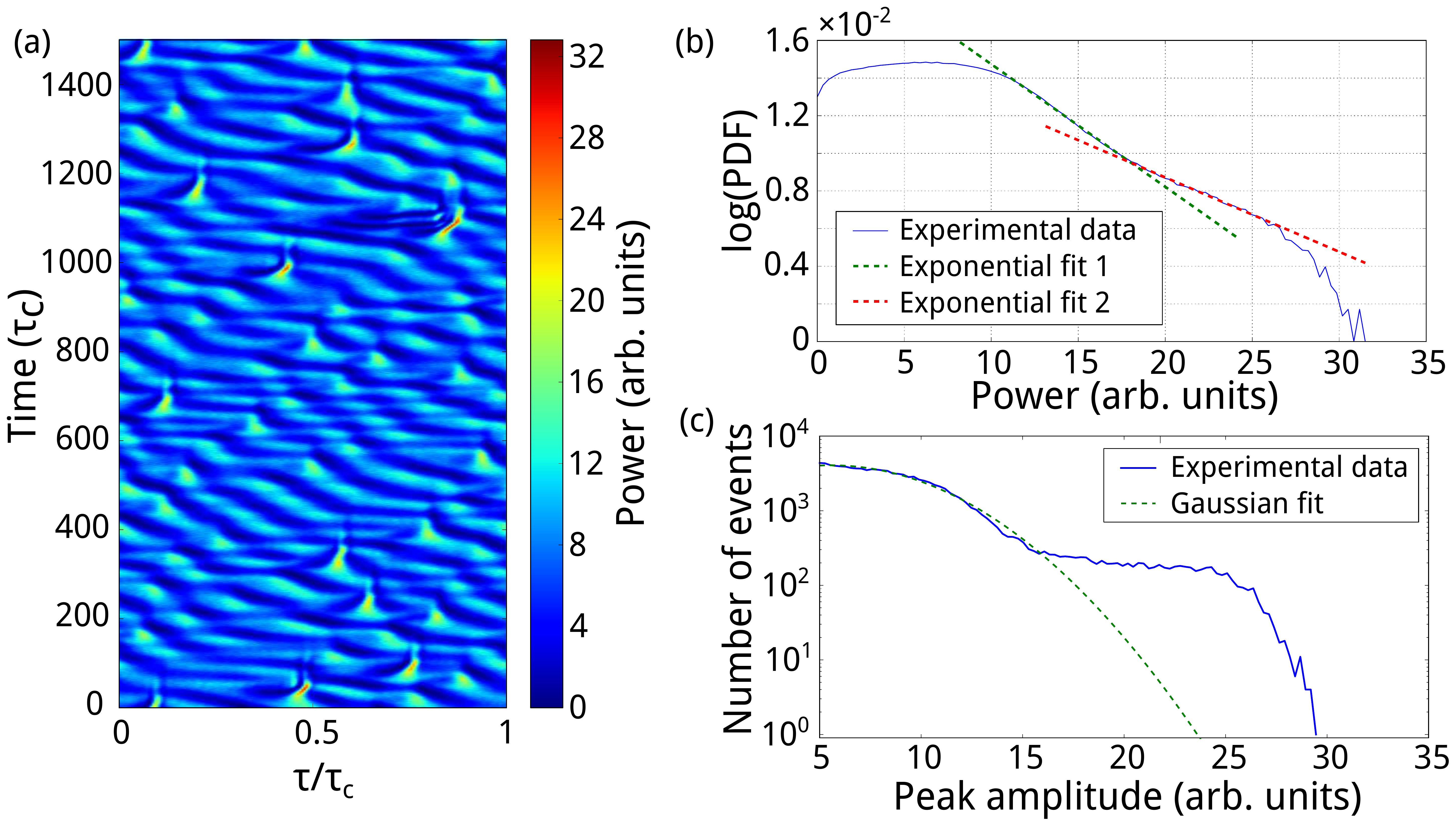}
\caption{Dynamical and statistical properties of extreme events. (a) Spatio-temporal diagram of the optical power along 1500 roundtrips, (b) logarithm of the probability density function of the total power and (c) distribution of the peak amplitudes. Both (b) and (c) are computed on long time traces of length $\simeq 27\, 000\, \times\,  \tau_c$ ($5 \times 10^6$ points)}
\label{xt_stats}
\end{figure}
\begin{figure}[htb]
\centering\includegraphics[width=.5\textwidth]{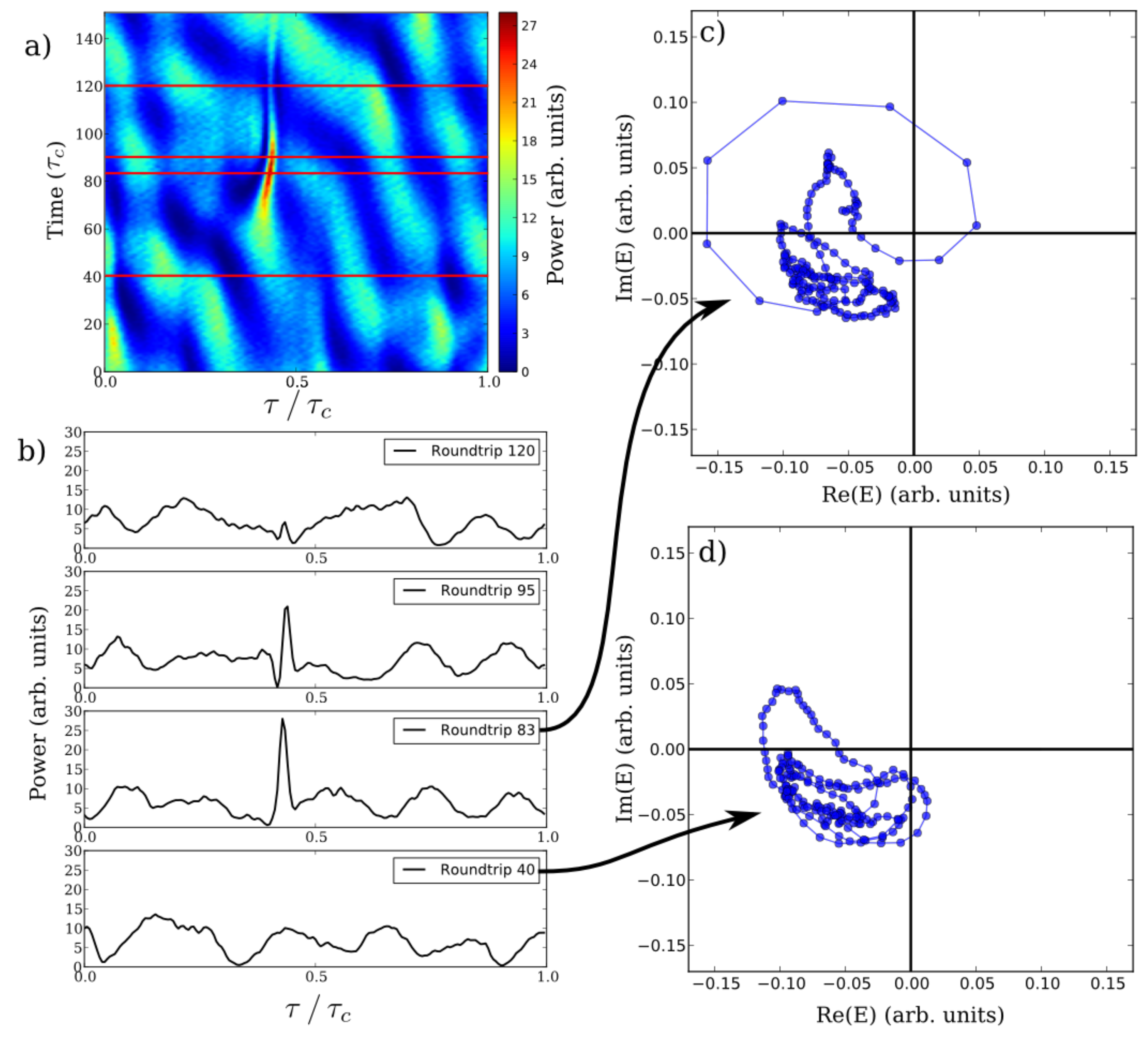}
\caption{Power and phase dynamics around the formation of an extreme event. (a) spatio-temporal diagram focused on one event, taken around roundtrip 1000 of Fig.~\ref{xt_stats}(a), (b) Power profiles corresponding to roundtrip 40, 83, 95 and 120 and depicted by red horizontal lines in (a), (c) phase trajectory in the Argand plane before the extreme event (roundtrip 40) and (d) at the maximum peak power of the event (roundtrip 83).}
\label{ext_event_full}
\end{figure}
\begin{figure}[ht]
\centering\includegraphics[width=.5\textwidth]{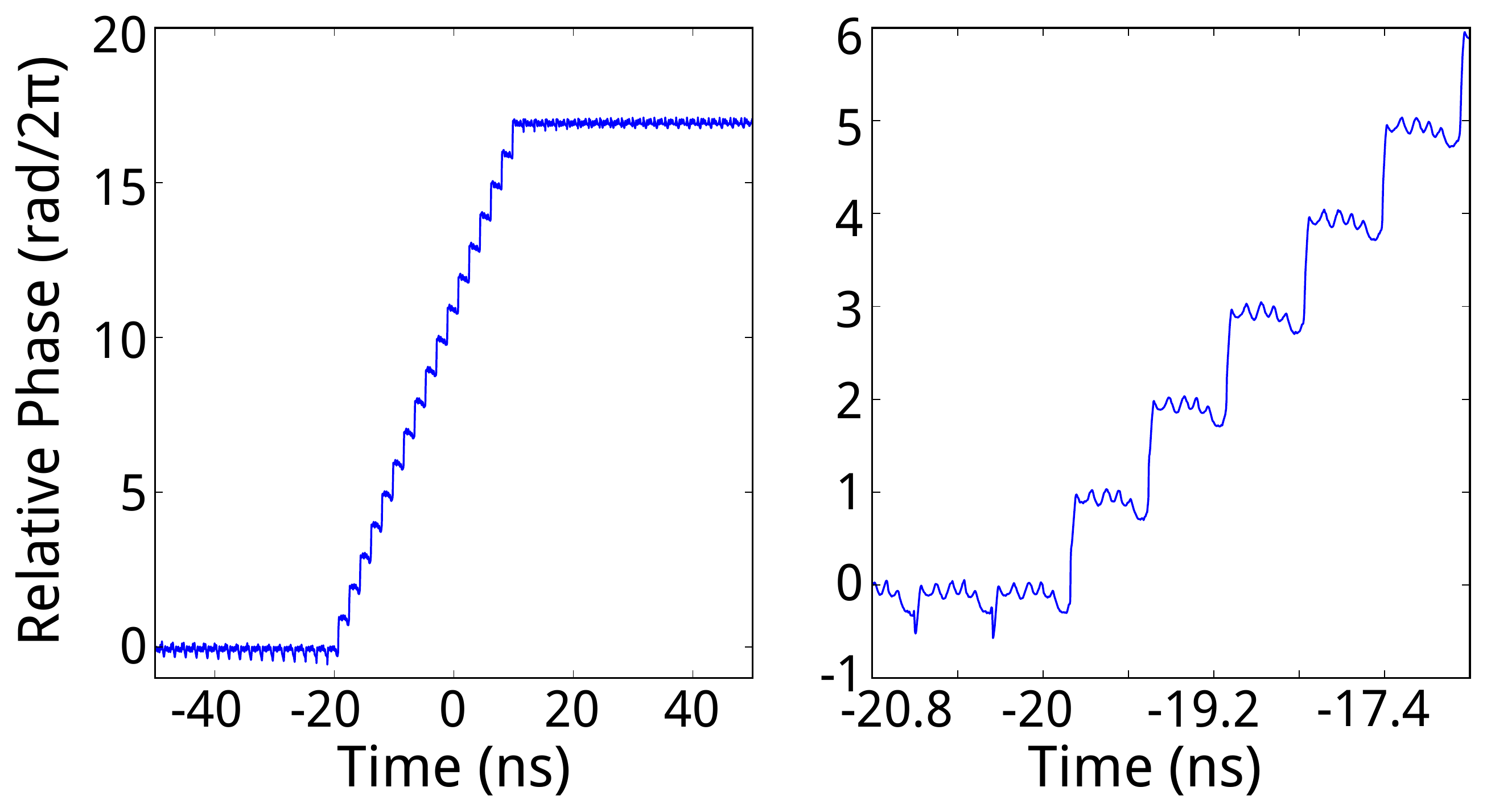}
\caption{Evolution of the phase when approaching the extreme event. Left : 100~ns long time trace of the phase centered on the maximum of an extreme event showing the acquisition of positive phase rotations. Right : zoom of the same time trace at the beginning of the change of slope. The initial phase is arbitrarily set to zero.}
\label{phase_steps}
\end{figure}
\begin{figure}[tb]
\centering
\includegraphics[width=10cm]{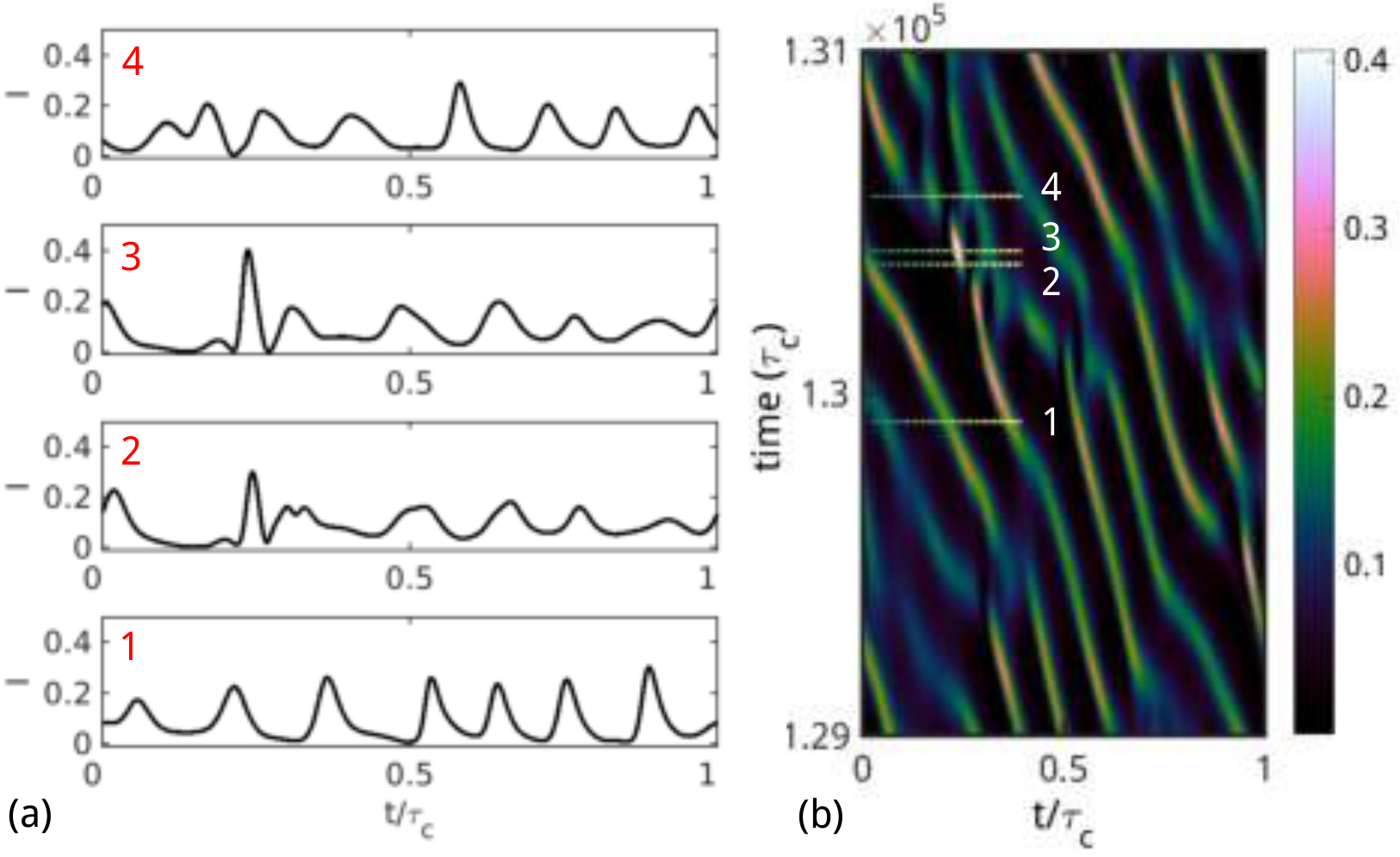}
\caption{Zoom (b) of the spatiotemporal diagram centered on an event of high intensity (3) and intensity time traces (a) at fixed roundtrip corresponding to the horizontal cuts highlighted on the diagram.}
\label{fig:spa_temp_sim}
\end{figure}
\begin{figure}[ht]
\centering
\includegraphics[width=12cm]{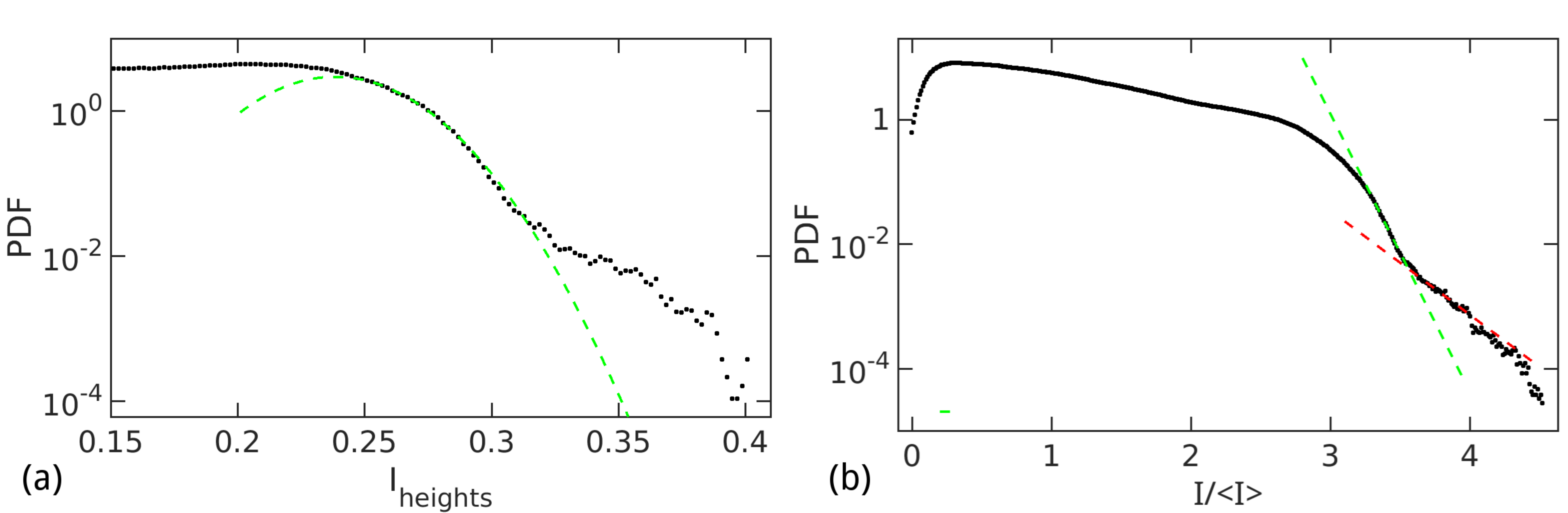}
\caption{PDF of the temporal peak heights, computed over 2.1$\times$10$^6$ roundtrips (a), the green dashed line corresponds to a Gaussian fit of the initial slope. PDF of all the values explored by the intensity (b), computed over 2.1$\times$10$^5$ roundtrips. The green and red dashed lines correspond to two negative exponential fits highlighting the change of the slopes in the tail of the distribution.}
\label{fig:statistics}
\end{figure}
\begin{figure}[ht]
\centering
\includegraphics[width=10cm]{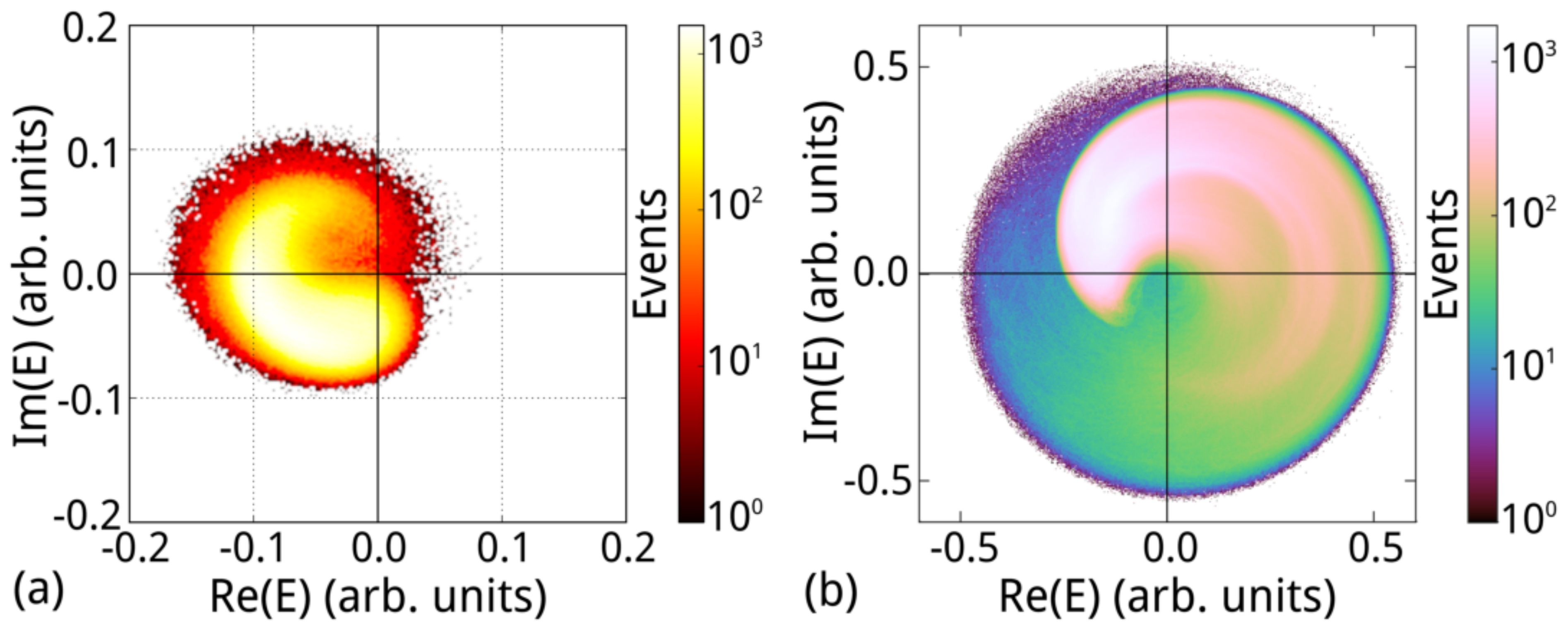}
\caption{Histogram of the values assumed by the electric field in the complex plane $(\textrm{Re}(E),\textrm{Im}(E))$ in the experimental case (a), computed over 10$^4$ roundtrips, and in the numerical case (b), computed over 10$^5$ roundtrips, in order to mantain the same number of events associated to the bounded state.}
\label{fig:phase_hist}
\end{figure}
\begin{figure}[ht]
\centering
\includegraphics[width=10cm]{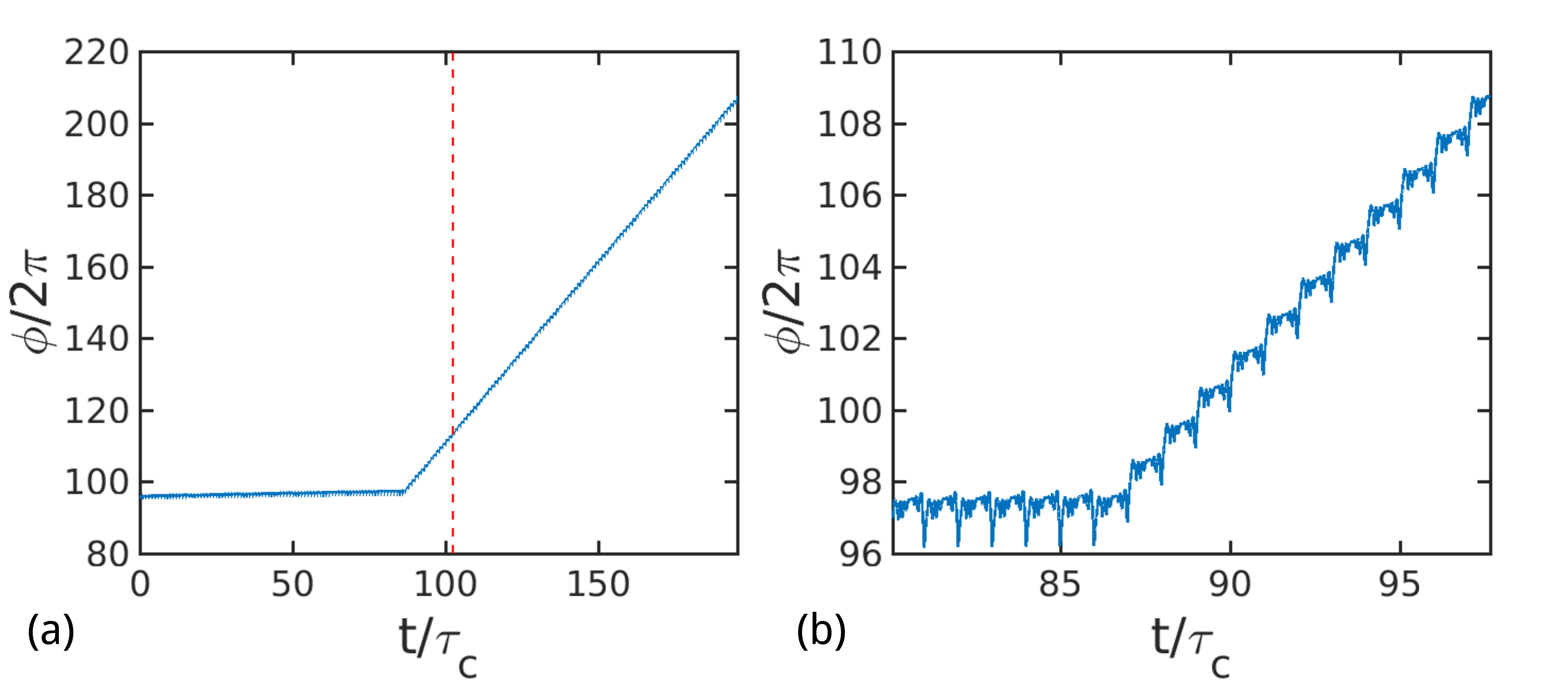}
\caption{Evolution of the phase when approaching an extreme event: the extreme event (occurring at the time indicated by the red vertical dashed line) is to be associated with a change in the slope of the phase. In particular in this case we notice an initially balanced situation between the number of positive and negative chiral charges that breaks when the negative chiral charge disappears. A little later an extreme event occurs.}
\label{fig:phase_jumps}
\end{figure}

Here %
$A$, $B$ and $C$ indicate the three fixed points for the chosen value of $y$ (vertical red dashed line): in particular, in the 0D model and in the subspace($\Re(E),\Im(E)$), $A$ is a node, $B$ is a saddle and $C$ is a focus \cite{gustave2015dissipative,gustave2016phase}. In the following, the phase--space trajectory of the system state will be studied, in relation to these three attractors.

In Fig. \ref{fig:stat_curve}(b) the stability scenery of the homogeneous stationary solution is plotted in the $(I,n)$ plane, where $n$ is the order of the longitudinal sidemode. The green domain indicates the unstable region in presence of diffusion ($d = 10^{-6}$), while the blue domain corresponds to the choice $d=0$, which amounts to the pure rate equation model without spectral filtering. It is clear that the diffusive term sets a limit to the number of unstable longitudinal modes but does not significantly affect the instability threshold and its local curvature.

The stability diagram of Fig. \ref{fig:stat_curve}(b) is typical of the condition $\theta+\alpha<0$, which means that the injected field is \textbf{blue} detuned with respect to the slave laser frequency. In this case a multimode instability affects a part of the upper branch close to the left turning point. The homogeneous stationary solution is unstable and a roll pattern arises due to the beating of the fundamental mode with high order sidemodes. Eventually even the rolls become unstable when the injection amplitude is brought too close to the left turning point, and more complex spatio-temporal dynamics can be observed, as in the experiments.

The experimental setup is based on a highly multimode semiconductor ring laser under coherent external forcing \cite{gustave2015dissipative, gustave2016phase,gustavethesis}. The ring laser is made of a semiconductor optical amplifier (SOA) acting as gain medium centered around $980$~nm and enclosed in a 1~m long ring cavity. The anti-reflection coating on the SOA facets prevents from self-lasing of the semiconductor element. An example of an experimental observation of such a dynamical regime in which an unstable roll pattern hosts high intensity peaks appearing randomly in space and time is shown in Fig.\ref{xt_stats}.

Fig.\ref{xt_stats}(a) shows a spatio-temporal diagram, where the horizontal axis refers to the longitudinal spatial dimension of the laser and the vertical axis represents time in units of the cavity roundtrip time $\tau_c \simeq 3.6$~ns. In this figure, we can clearly identify the unstable roll pattern in background and high intensity pulses nucleated on top of it. In order to get the statistical signatures of this regime, we computed the probability density function (PDF) of the total optical power and the distribution of the peak amplitudes, respectively represented in \ref{xt_stats}(b) and \ref{xt_stats}(c). 

In the PDF of the power, we first distinguish a plateau (up to 10), followed by an exponential decay (up to 17) and a second exponential decay with smaller slope corresponding to heavy tail statistics. The presence of this statistical deviation at high amplitudes implies that high intensity pulses occur more often than what expected looking at the first exponential decay. %

In the PDF of the peak heights shown in \ref{xt_stats}(c) the heights are the difference between each maximum and the previous minimum of the full time trace. %
The shape of the distribution is well fitted by a Gaussian up to 16 and presents a huge deviation beyond that point. The Gaussian part of the distribution (at low values) is related to the oscillating background whereas the deviation appearing at high values corresponds to the presence of extreme events.

To understand the formation of these extreme events, we focus on one particular event, statistically representative of the relevant regime, and provide a detailed analysis by looking at both phase and intensity dynamics. Fig.~\ref{ext_event_full}(a) shows a zoom of the spatio-temporal diagram presented in Fig.~\ref{xt_stats}(a) on 150 roundtrips around an extreme event. The horizontal red lines represent the positions at which intensity profiles shown in Fig.~\ref{ext_event_full}(b) have been taken. The first line (roundtrip 40) shows the state of the system before the extreme event occurs and the associated phase trajectory is presented in Fig.~\ref{ext_event_full}(d). At this stage of evolution, the phase remains in a bounded region of the complex plane (Argand plane).

At roundtrip 83, the extreme event has reached its maximum intensity. By comparing the height of this event with the statistics shown in Fig.~\ref{xt_stats}(c), we can easily understand that this object participates to the strong statistical deviation described before.
As we can see in Fig.~\ref{ext_event_full}(c), the high intensity peak is accompanied by a striking topological change of the phase space picture, because now the trajectory circles around the origin. The system seems to be pushed away from the bounded state and to explore a wider region of the complex plane. By crossing the origin of the complex plane a defect is created and the extreme event acquires a counterclockwise ($+2\pi$) phase rotation. 

In Fig.~\ref{phase_steps}, we show the evolution of the unfolded phase around the formation of an extreme event. The time trace is centered on the extreme event maximum intensity ($t=0$). In this representation, it is clear that the phase stays bounded up to 20~ns before the maximum of the extreme event. We can clearly identify the phase rotations as a change of slope, that starts before the moment where extreme event reaches its maximum. At each roundtrip, the system acquires $2\pi$ rotations corresponding in this representation to steps of $2\pi$. The phase and amplitude dynamics leading to the emergence of these extreme events are reminiscent of the defect mediated turbulence discussed theoretically in the context of forced oscillatory media \cite{coullet1989defect,Gibson16}.

The experimental findings are well reproduced by numerical simulations made with the parameters of Fig. \ref{fig:stat_curve} and $y=0.004$.

In Fig. \ref{fig:spa_temp_sim} we show a zoom of a typical spatiotemporal regime when the rolls pattern is unstable. The diagram is centered on an event of high intensity occurring at the point highlighted by line 3.
The four temporal profiles illustrated in Fig. \ref{fig:spa_temp_sim} (a) correspond to the horizontal cuts highlighted in Fig. \ref{fig:spa_temp_sim} (b). In particular the temporal profile from section 1 occurs 500 roundtrips before the event and corresponds to a mostly bounded regime, where we can appreciate the remnants of the roll pattern which would be stable for a larger injection amplitude. Section 3 is taken at the maximum of the high intensity event, while sections 2 and 4 are taken, respectively, 40 roundtrips before the event and 160 roundtrips after.

For the statistical analysis, we plot in Fig. \ref{fig:statistics} the PDF of the temporal peak heights (maximum minus precedent minimum) (a) and of the overall intensity values (b). Similarly to the corresponding experimental Fig. \ref{xt_stats}, we observe that the PDF of the heights (a) presents an initial slope that follows a Gaussian distribution (green dashed line), to be associated with the bounded state, while the events in the tail of the distribution appear more frequently than expected by a Gaussian statistics. Furthermore, in the statistics of the total intensity we notice two different negative exponential slopes in the tail of the distribution, as observed in the experiments: the first slope (green dashed line fit) may be associated with the roll pattern, the second slope (red dashed line fit) highlights the heavy tail where very intense events are more frequent than expected in the bounded regime.

Let us focus now on the role of the phase dynamics in the formation of these extreme events.
In order to compare with the experiment, we computed a histogram in the complex plane $(\textrm{Re}(E),\textrm{Im}(E))$ of all the values assumed by the electric field. Fig. \ref{fig:phase_hist}(a) is the experimental histogram and Fig. \ref{fig:phase_hist} (b) is the numerical one. A clear similarity between the two histograms is noticeable, especially regarding the presence of an unstable bounded state (yellow area in (a), pink-white area in (b)) from which rarer events get detached, exploring other areas of the histogram (red area in (a), blue-green area in (b)).

From the experiments it was already clear that chirality was playing an important role in the formation of extreme events: in particular the latter seem to be associated with the appearance of a positive (counterclockwise) chiral charge (see Fig. \ref{phase_steps}). In the numerical simulations, we noticed that clockwise phase rotations also play an important role in the formation of extreme events as well.

In Fig. \ref{fig:phase_jumps} (a) we plotted the evolution of the phase when approaching an extreme event, in a situation similar to the experimental case: here we notice that an initial balance between the number of positive and negative chiral charges is broken by the loss of a negative charge, which gives rise to a change in the slope of the phase. A little later, an extreme event occurs (highlighted by the red vertical dashed line. In the zoom in Fig. \ref{fig:phase_jumps} (b) it is possible to notice that in the initial plateau for the phase there are both a positive and a negative chiral charge, which then get reduced to a single positive charge.

We can argue that the occurrence of extreme events may be associated with a global change in the slope of the phase: however, as it has been observed in the experiment, this change is not to be considered as a precursor since it can occur either before or after the extreme event itself.

\begin{figure}[tb]
\centering
\includegraphics[scale=0.4]{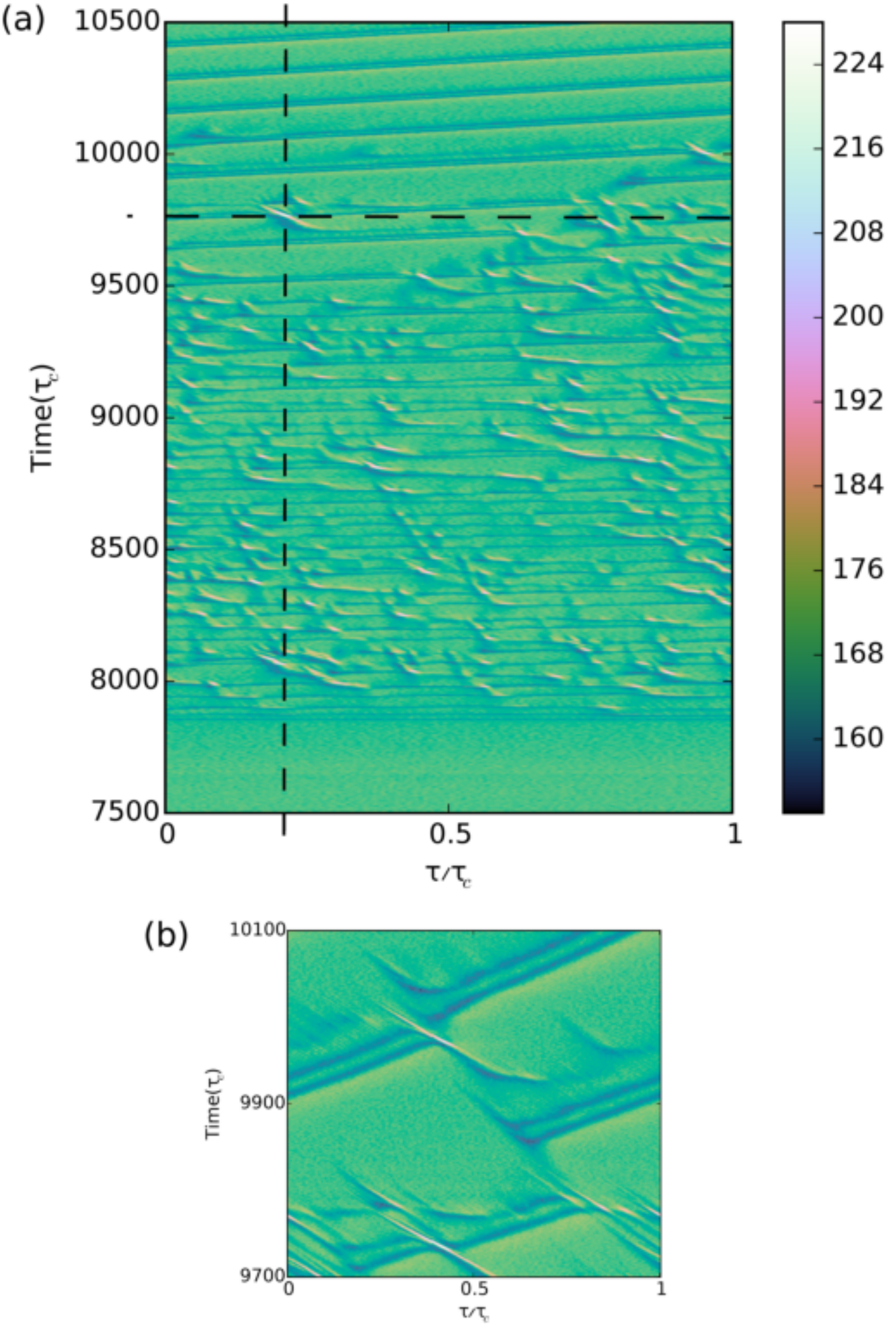}
\caption{(a) Space time diagram showing the different regimes observed in the experiment. Collisions between dissipative phase solitons and counter-propagating structures lead to the emergence of extreme events. Before 7500 the system is stably locked and after 10500 one soliton is stable. (b) Observation of collision in a slightly modified reference frame (6450~ps roundtrip time instead of 6470).  The laser threshold current is 18 mA, here its bias current is set at 1.6 times above the threshold.}
\label{fig2:diag_spatio_temp}
\end{figure}
\begin{figure}[htb]
\centering
\includegraphics[scale=0.3]{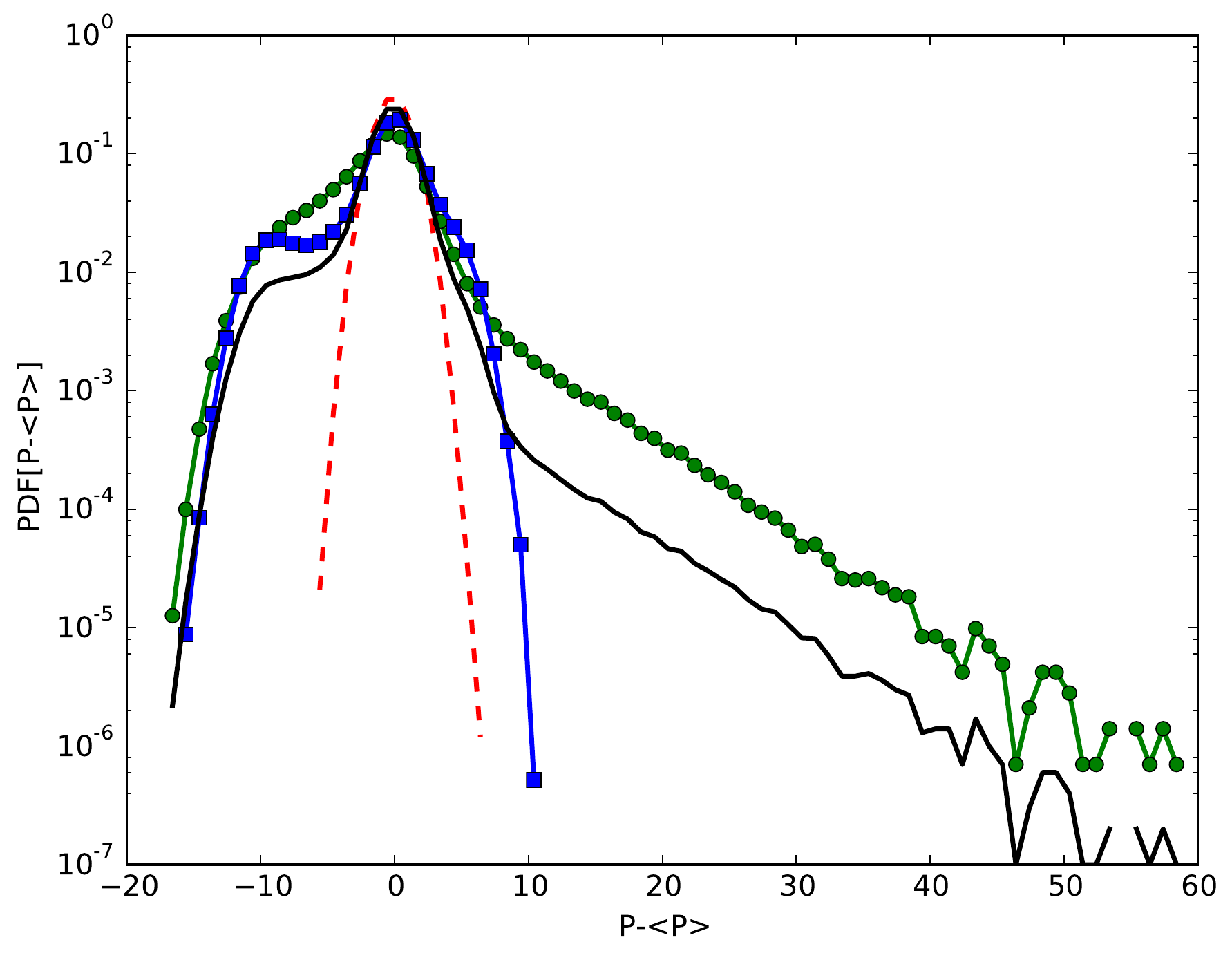}
\caption{Experiment: Probability Density Function (PDF) of the power fluctuations along the recording time trace. The PDF shows the deformation of the statistics with the emergence of localized events with high amplitude. Dashed red line : PDF from 0 to 7751 (locked state); Green line with circles: from 7850 to 10060 (collision regime); Blue line with squares:  from 10500 to 15500 (stable soliton propagation); Black line: PDF from 0 to 15500.}
\label{fig3:stat_fig17}
\end{figure}
\begin{figure}[htb]
\centering
\includegraphics[width=10cm]{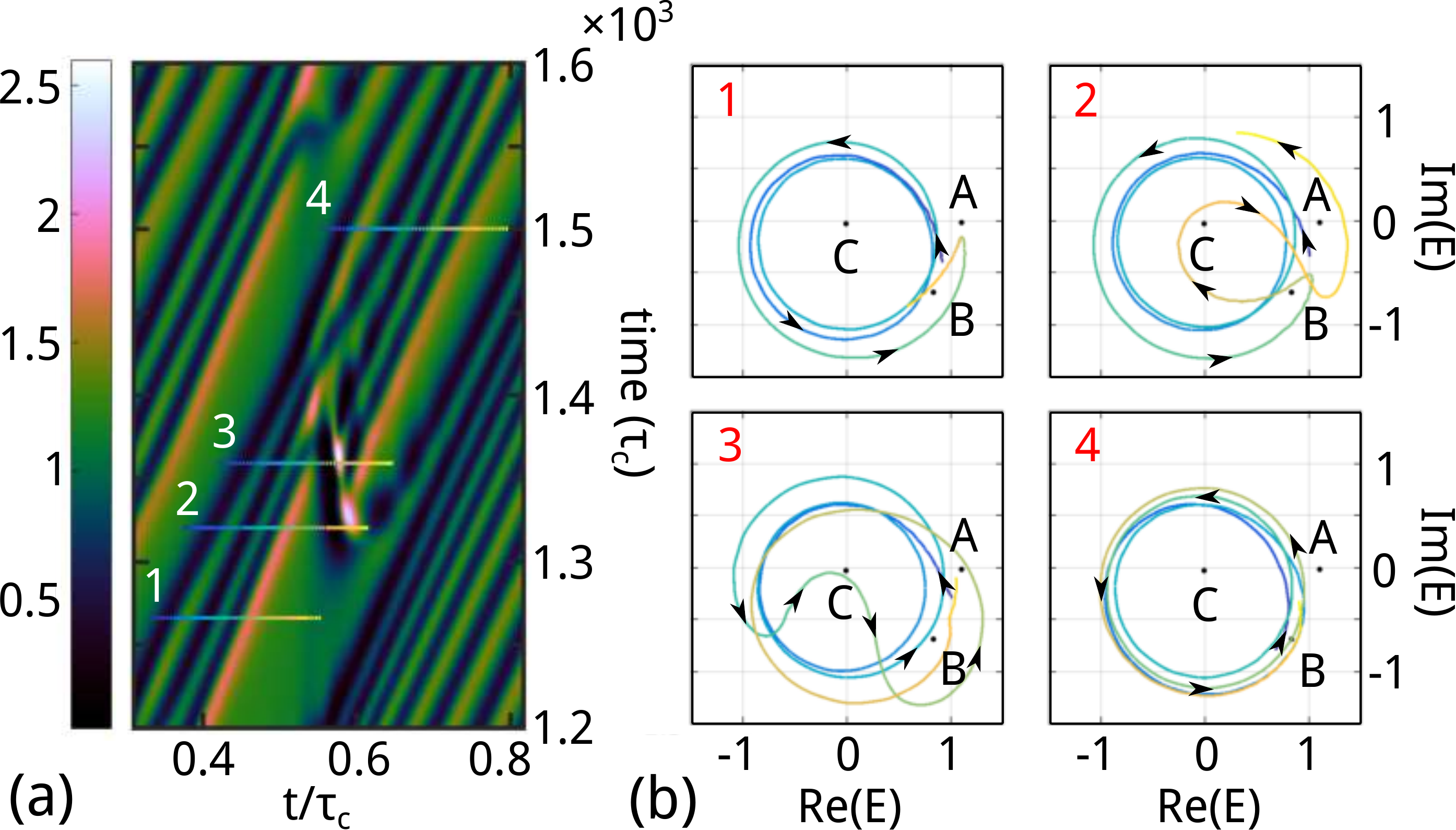}
\caption{Numerical simulation: zoom of the spatiotemporal diagram of the electric field intensity around a collision (a), and trajectories in the Argand plane for the selected roundtrip sections (b). Roundtrip section 3 coincides with the occurrence of a high intensity event and it is preceded by a clockwise rotation at the roundtrip section 2.}
\label{fig5:numerics}
\end{figure}

We now turn our attention to a slightly different experimental setup where a Fabry-Perot resonator is used instead of a ring \cite{gustave2017formation}. Now the active medium is antireflection coated only on one side. A lens is inserted to focus the output field on a high reflected mirror (99\%) which closes the one meter-long cavity. Two 10\% beam splitters are inserted inside the Fabry-Perot resonator to provide both an input for the forcing field and an output for the detection of the emitted field. The master laser is provided by a grating tunable external cavity semiconductor laser. In order to prevent all reflections from the slave laser to the master laser, an optical isolator is inserted. %

In this configuration the formation of phase soliton complexes was observed \cite{gustave2017formation}: two phase solitons with one chiral charge attract and collide to form a single phase soliton with a double chiral charge, whose velocity is different from that of the single-charge phase soliton, and the process can repeat leading to complexes with high chiral charge.

As outlined in the Introduction, we hypothesized that phase soliton collisions could be linked to emergence of extreme events, so we focused on new regimes where an underlying irregular dynamics could couple with the propagation of self-organized structures. Here we have selected a specific regime where we are able to observe the emergence of a spatiotemporally localized event with a high amplitude.

Fig.~\ref{fig2:diag_spatio_temp}(a) represents a spatiotemporal diagram of such a regime. We can observe mainly three different behaviors. During the 7751 first roundtrips, the system is in synchronized state. Between roundtrips 7850 and 10060, the system exhibits a markedly different and complex dynamics. We can recognize the propagation of dissipative phase solitons, often with double charge. These coherent structures, which are spontaneously nucleated from the synchronized state, propagate and sometimes collide with different structures propagating in the opposite direction in this reference frame %
, inducing extreme events localized in time and space.%

In Fig.~\ref{fig2:diag_spatio_temp}(b) 
we see that before the collision, the dissipative phase soliton with chiral charge two collides with a different particular structure. After this collision, one hump of the soliton complex seems to be annihilated and a new hump is regenerated about ten roundtrips later. This double structure propagates and eventually collides again with a counterpropagating structure, giving rise to a new extreme event. Finally, after this complex regime (from roundtrip number 10400 to the end of this record), a regular regime is attained, dissipative phase solitons with a chiral charge of two are stable and propagate without collisions.

The statistical impact of these events can be characterized by the computation of the PDF of the power fluctuations for a long time trace as it is represented in Fig.~\ref{fig3:stat_fig17}. The black solid line represents the PDF on all the roundtrips, which is asymmetric and shows a clear heavy tail at the highest power values. The dashed red line, green line with circles and blue line with squares correspond respectively to the synchronized state, the turbulent regime and the soliton propagation regime described in Fig.~\ref{fig2:diag_spatio_temp}(a). In the synchronized state, where the emission is regular and continuous, the PDF results essentially from the physical and detection noise (dominant), which is stochastic. This explains why the PDF has a gaussian form. In the opposite case, in the turbulent regime, the emergence of extreme events modifies strongly the PDF which deviates from the gaussian statistics to exhibit a heavy tail statistics. Finally, the propagation of dissipative phase solitons plays also a significant role in the modification of the PDF. Due to their particular shape including a depression on the trailing edge \cite{gustave2015dissipative,gustave2016phase}, the probability of the low power values is increased.

We performed numerical simulations in the same regime observed in the experiment, that is, where the injected field is red detuned with respect to the slave laser frequency ($\theta+\alpha>0$). Precisely, we set $\theta=-2.7$ and assumed that the laser is well above threshold $\mu=2$, as in the experiment, with injected amplitude $y=0.11$. The remaining parameters are the same as in the previous simulations.
The stability scenery with these parameters differs from that of Fig. \ref{fig:stat_curve}, because the whole upper branch of the homogeneous stationary solution now is stable and the system can sustain phase solitons (PSs), provided the amplitude of the injected field is such that the fixed points $A$ and $B$ described above are sufficiently close, i.e. the system is close to the left turning point of the homogeneous stationary solution \cite{gustave2015dissipative,gustave2016phase,gustave2017formation}. Phase solitons are homoclinic trajectories connecting the locked homogeneous state $A$ with itself, through a rotation of the phase of electric field by (multiples of) $2\pi$. Therefore, phase solitons posses a chiral charge, and in a non-instantaneous propagative (temporal) system as ours, the chiral charge can only be positive (\textit{i.e.} the phase rotates counterclockwise) \cite{longhi1998nonlinear,gustave2016phase}.

In Fig. \ref{fig5:numerics}(b) we show a zoom of the spatio-temporal diagram in a simulation where the initial condition was the homogeneous stationary solution with a superimposed phase kink of 4$\pi$ (but the same kind of regime can develop starting from noise). We can notice that the spatiotemporal dynamics present some clear similarities with the experimental data. In particular we can observe the presence of phase solitons and the emergence of some peculiar structures moving at a different velocity that eventually collide with the phase solitons, giving rise to an event of high intensity.

In Fig. \ref{fig5:numerics}(a) we show the phase diagrams relative to the horizontal cuts highlighted in Fig. \ref{fig5:numerics}(b), time grows from red to yellow. The points $A$, $B$ and $C$ are the projections in the Argand plane of the three fixed points \cite{gustave2015dissipative,gustave2016phase}, respectively a node (stable for our choice of parameters), a saddle and an unstable focus. In the first frame, relative to roundtrip section 1, we can observe a phase soliton complex \cite{gustave2017formation} of charge 3 that propagates inside the cavity: the trajectory of the system consists in three  counterclockwise phase rotations in the Argand plane, passing only once close to point $A$. At roundtrip section 2 a new object emerges at the right side of the soliton complex, together with a clockwise phase rotation (charge -1) in the Argand plane. The interaction between the two structures gives rise to a high intensity event at roundtrip section 3, where the clockwise rotation has already been lost. The interaction has also the effect of altering the phase soliton complex velocity inside the cavity, as can be noticed just after roundtrip section 3. At roundtrip section 4 the complex has regained its shape, with one additional charge, coming from the interaction.

The above evidences indicate that the appearance of these short-lived pulses with clockwise phase rotations that collide with stable phase solitons is the basic physical mechanism responsible for the extreme events observed in this system.

\section{2D}\label{sec:2d}In the previous sections, we have seen excellent examples of the generation of extreme events due to external forcing in systems described by ordinary differential equations (Section \ref{sec:0d}) or by partial differential equations in time and one auxiliary dimension (Section \ref{sec:1d}). Here we describe the generation of rogue waves due to external forcing in the dynamics of 2D optical systems with gain. In particular, we investigate spatio-temporal turbulence due to interacting optical vortices in lasers and in optical parametric oscillators (OPOs) with injection \cite{Gibson16}. Transverse 2D rogue waves have also been found in lasers with saturable absorbers \cite{Selmi16,Rimoldi17} and even in passive optical systems in the presence of pattern competition \cite{Eslami17}. The focus of this section is, however, on the specific effect of the external forcing via an optical injection on inducing 2D rogue waves (RWs), i.e. high intensity peaks with non-Gaussian statistics, in the transverse section of laser systems.

We investigate the spatio-temporal dynamics of two closely related (in terms of equations and gain threshold) optical systems. A class A laser with injected signal \cite{Mayol02} and a singly resonant optical parametric oscillator (SROPO) with seeding \cite{Oppo13}. The first case is closely related to systems described in the previous sections, the main differences being that in class A lasers such as dye, quantum dots, quantum cascade or He-Ne lasers, all material variables (for example energy level populations and material polarization) evolve on temporal scales much shorter than that of the cavity field. The SROPO case corresponds to parametric down conversion of a pump field into a signal and an idler field, whose frequencies sum to that of the pump, inside an optical cavity. In the singly resonant case, the frequencies of the signal and idler fields are very different and only the signal field is resonated inside the cavity. In the SROPO case the material variables dynamics is instantaneous since parametric down conversion utilises virtual energy levels that exist for times smaller than those given by the Heisenberg time-energy uncertainty principle.
In both cases we assume the external forcing takes the form of an optical injection detuned from one of the optical cavity resonances.

For a class-A laser with injection the model equations are
\begin{eqnarray}
\label{laser}
\frac{\partial E}{\partial t} &=& E_{IN}-(1-i\omega)\;E +  P \; \left(1-\frac{1}{3}|E|^2 \right) \; E + i\nabla^2 E 
- \Gamma (\omega+\epsilon \nabla^2)^2 E \, ,
\end{eqnarray}
while for the SROPO case with injection we have
\begin{eqnarray}
\label{sropo}
\frac{\partial E}{\partial t} &=& E_{IN}-(1-i\omega)\;E +  P \; \mathrm{sinc}^2(|E|)  \; E + i\nabla^2 E 
- \Gamma (\omega+\epsilon \nabla^2)^2 E \, ,
\end{eqnarray}
where $E$ is the complex field inside the cavity, $E_{IN}$ is the (real) amplitude of the external forcing, $\omega$ is the frequency difference between the unperturbed field and the external driver, $\nabla^2$ is the transverse Laplacian that describes diffraction, $P$ is the pump, $\epsilon$ is a small parameter (here fixed at $0.01$),  $\Gamma$ is equal to one for the Swift-Hohenberg correction \cite{Lega94} and zero otherwise. Eq. (\ref{sropo}) reduces to Eq. (\ref{laser}) when the pump $P$ is close to its threshold value of unity. We assume flat mirrors but note that in the case that the mirrors are curved, these equations can be projected onto a set of Laguerre-Gauss modes \cite{Dalessandro92} thus transforming partial differential equations into a large set of ordinary differential equations. Superposition of Laguerre-Gauss modes can, of course, generate interacting vortices in two spatial dimensions and several of the results presented below may find confirmation in these mode expanded equations.
\begin{figure}[t]
\centering
\includegraphics[width=0.8\linewidth]{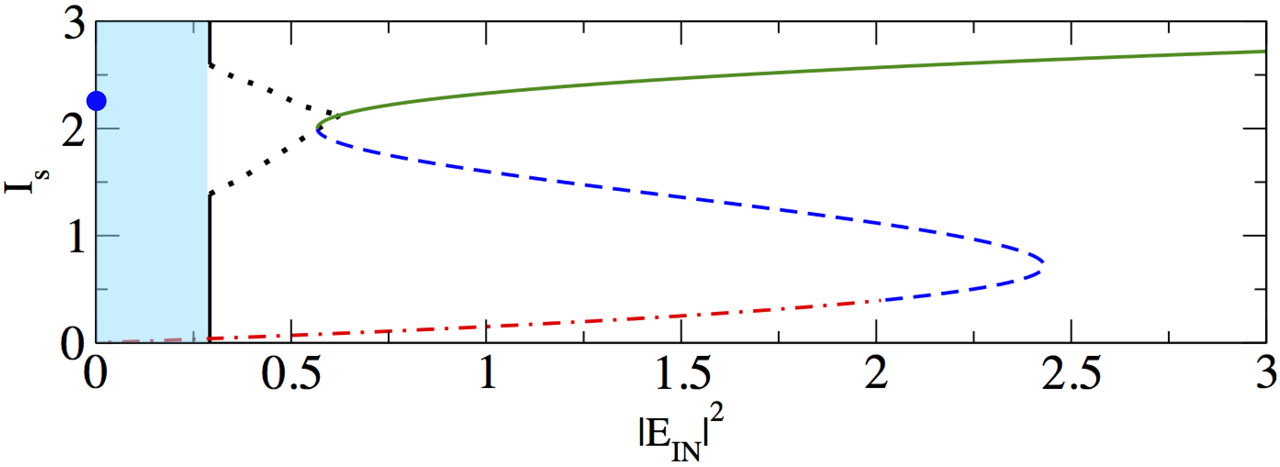}
\caption{Stationary intensity $I_s$ of the homogeneous solutions and their stability (stable $=$ solid green, unstable with real eigenvalues $=$ dashed blue, unstable with complex eigenvalues $=$ dash-dotted red) for the model (\ref{laser}) with $P=4$, $\omega=0.53$ and $\Gamma=1$. The black dotted lines represent the minima and maxima of stationary hexagonal patterns and the vertical lines where the optical turbulent state starts (shaded area). The blue circle is the stationary intensity of the laser with no injection.}
\label{fig1}
\end{figure}
We start by considering the spatially homogeneous stationary states of the two model equations. There are three parameters that can be controlled experimentally: the detuning between slave laser and injection $\omega$, the amplitude of the injection $E_{IN}$ and the external pump $P$. For fixed values of $\omega$ and $P$, the stationary intensity $I_S$ of the homogeneous states has a typical $S$-shaped dependence on $E_{IN}^2$ as displayed, for example, in Fig. \ref{fig1}. By increasing (decreasing) the pump value $P$, the $S$-shaped curve stretches upward (downward) while keeping its intersection with the origin fixed. The uppermost lines in the $S$-shaped curves of Fig. \ref{fig1} correspond to the homogeneous locked states where the external driving is large enough to overcome the frequency difference with the injected device $\omega$. When increasing $E_{IN}$, a saddle-node bifurcation heralds the onset of the frequency and phase locked homogeneous states. In the [$|E_{IN}|^2;\;I_s$] parameter space, the coordinate of the turning point where the saddle-node bifurcation takes place is well approximated by [$\omega^2\;(P-1); \;P-1$] \cite{Oppo86}. When, instead, decreasing the parameter $E_{IN}$, the homogeneous solution loses stability to spatially periodic patterns with a critical wave-vector given by $k_c=\sqrt{\omega}$. In Fig. \ref{fig1} the maximum and minimum intensities of the hexagonal patterns obtained numerically when reducing the external driver are displayed via a black dotted line. Although the phase of the pattern is periodically modulated in space, the stationary character of these pattern solutions demonstrates that they are locked to the frequency of the injection. 

\begin{figure}[t]
\includegraphics[width=0.5\linewidth]{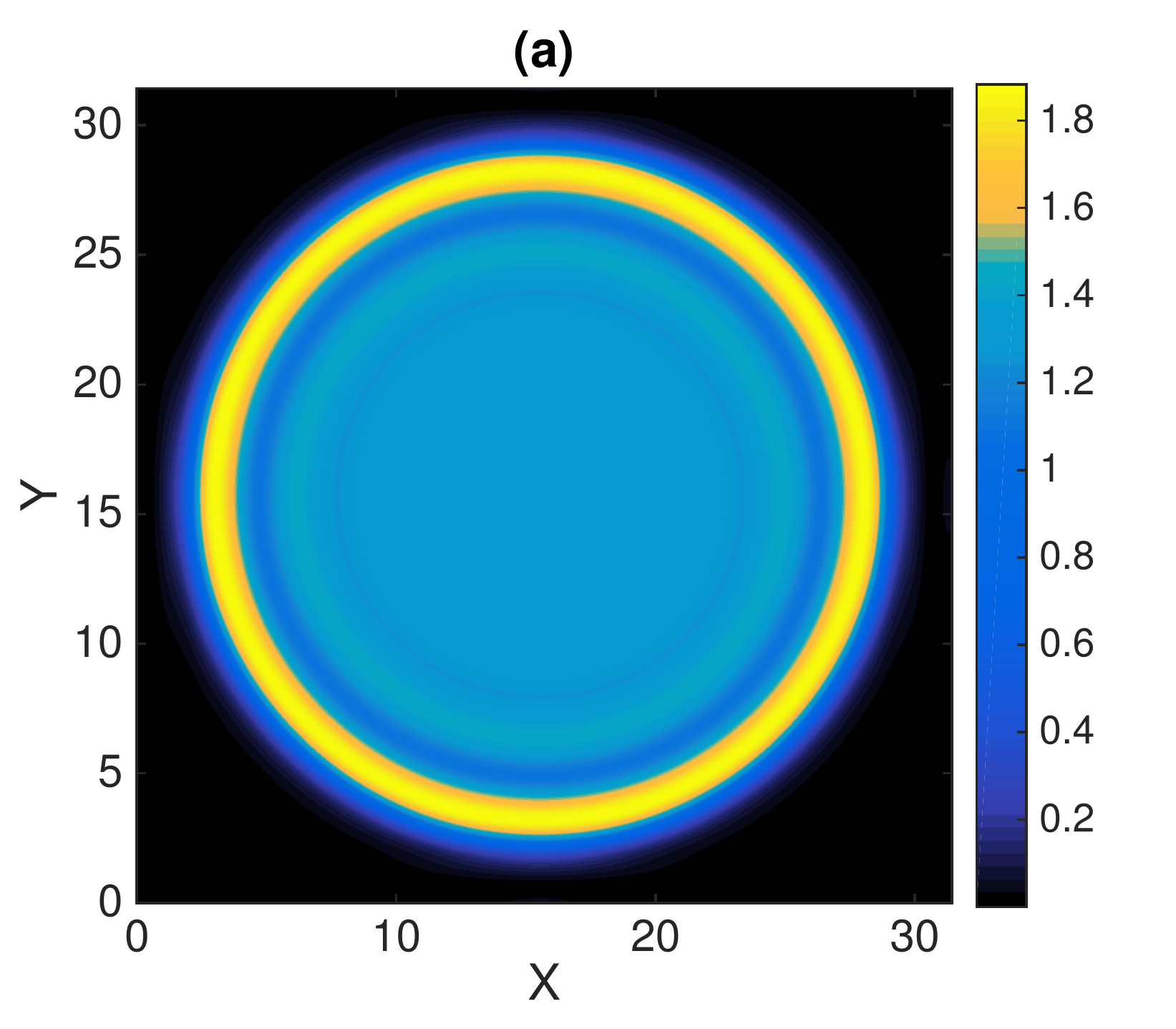}
\includegraphics[width=0.5\linewidth]{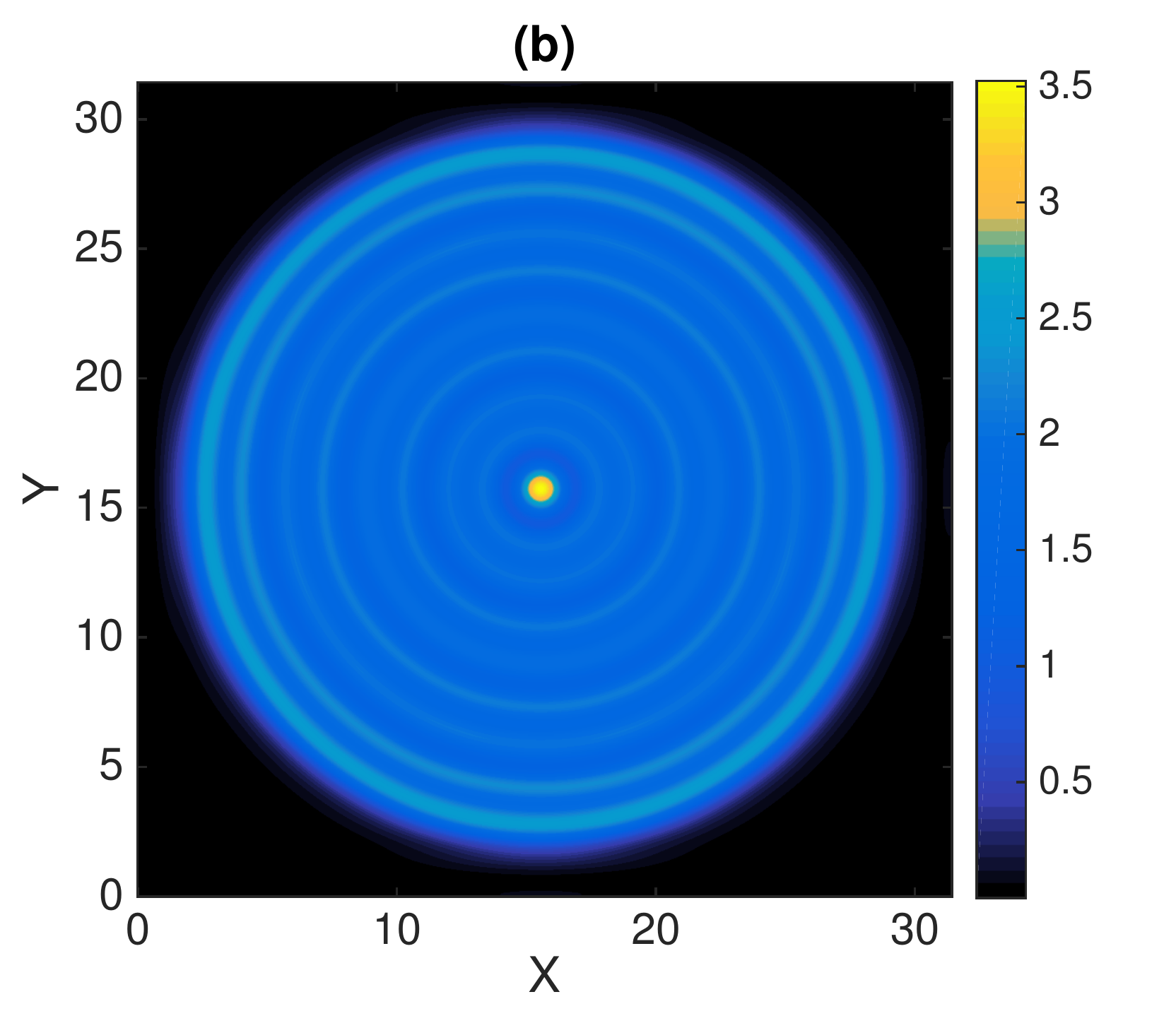}
\includegraphics[width=0.5\linewidth]{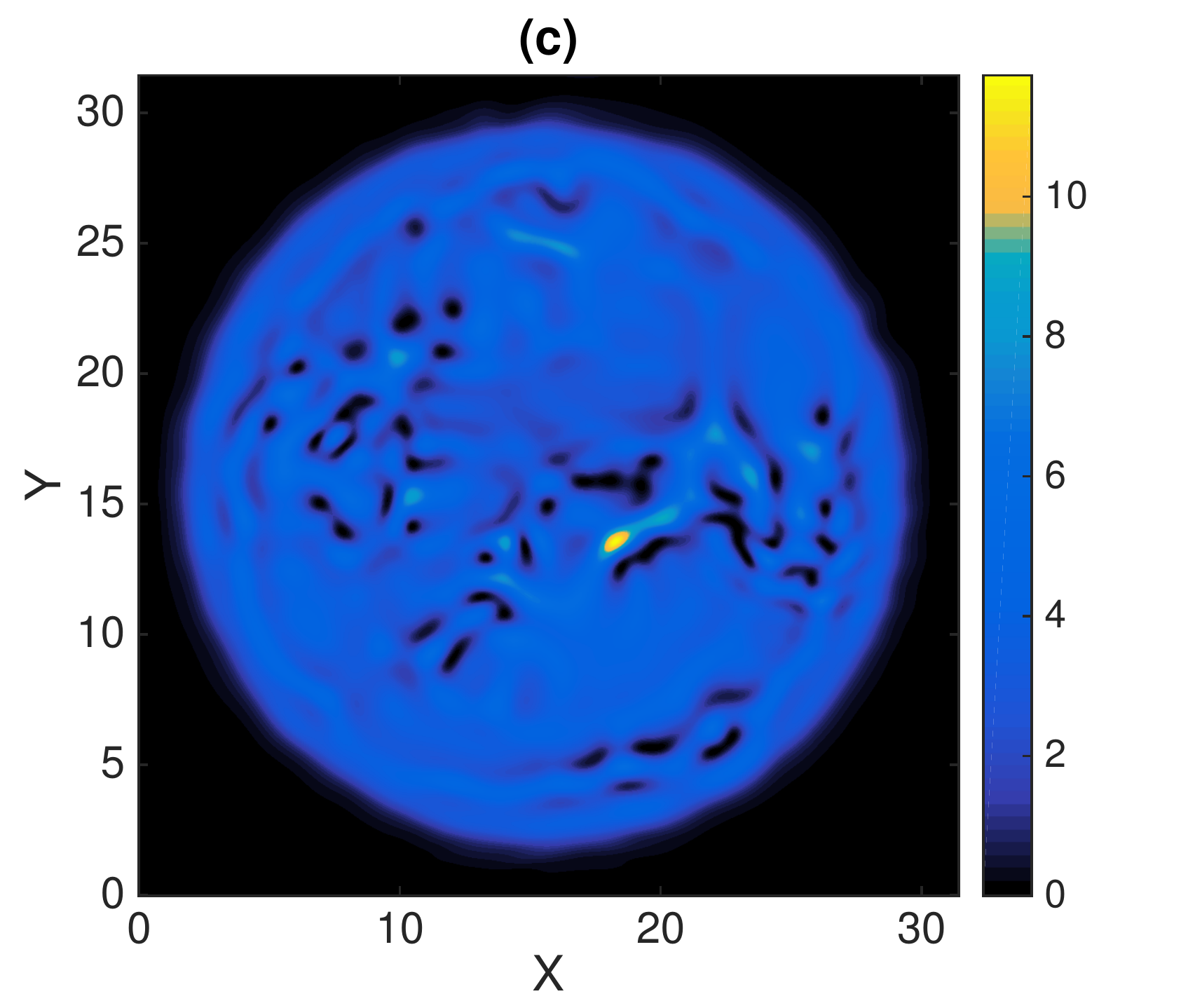}
\includegraphics[width=0.5\linewidth]{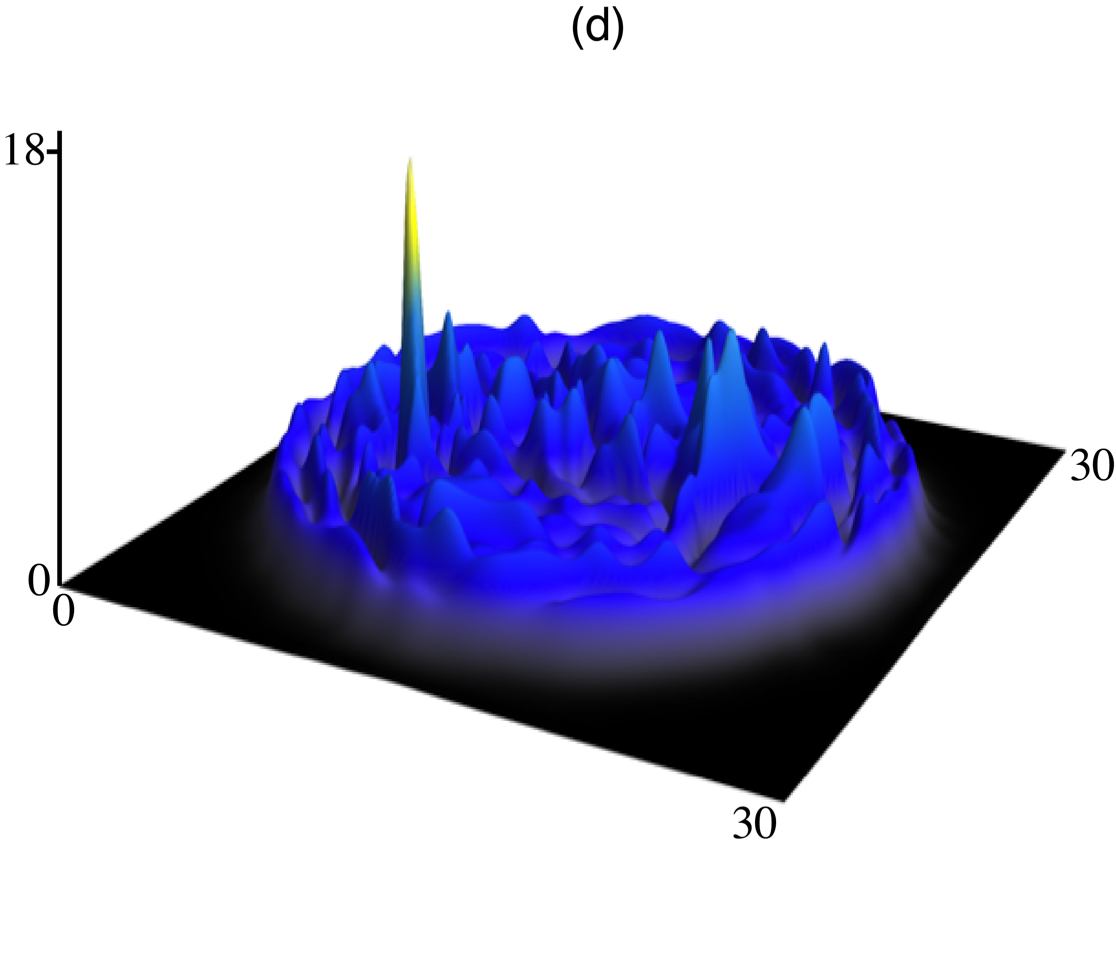}
\caption{Transverse intensity profiles for the SROPO model (\ref{sropo}) with $\omega=1$, $E_{IN}=1.2$ and $\Gamma=0$. (a) Close-to-threshold stationary state for $P=1.1$; (b) Modulated pattern for $P=1.96$; (c)-(d) optical vortex-mediated turbulence for $P=4$. Panel (d) shows a 3D representation of the turbulent instensity displaying a very large intensity peak.}
\label{fig2}
\end{figure}

By further decreasing the injection amplitude $E_{IN}$, one enters into a region of spatio-temporal chaos \cite{Gibson16}, i.e. the regime of optical turbulence of relevance here. By using the input pump $P$ as the control parameter while keeping the detuning $\omega$ and the injection amplitude $E_{IN}$ fixed, we demonstrate here that the transition to a turbulent state dominated by interacting vortices is a generic property of both laser and OPO models. 
Just above the threshold for laser or OPO signal generation, i.e. $P$ above but close to the value $1$, the system output is locked to the injection and is basically homogeneous apart from finite size diffraction rings (see Fig.~\ref{fig2} (a)). By increasing $P$ an instability to a stable but periodically modulated structure with a strong central peak sets in (see Fig.~\ref{fig2} (b)). Such spatial structure is strongly affected by the transverse boundary conditions that, in this case, are due to the finite size of both pump and injection beams. By further increasing the pump $P$, a symmetry breaking bifurcation takes place and the system moves into a turbulent state (see Fig.~\ref{fig2} (c)) characterised by regions where the intensity is close to zero and others corresponding to high intensity peaks  (see Fig.~\ref{fig2} (d)). Note that the intensity peaks appear at random positions in the transverse plane during the spatio-temporal evolution.

\begin{figure}[htb]
\includegraphics[width=0.5\linewidth]{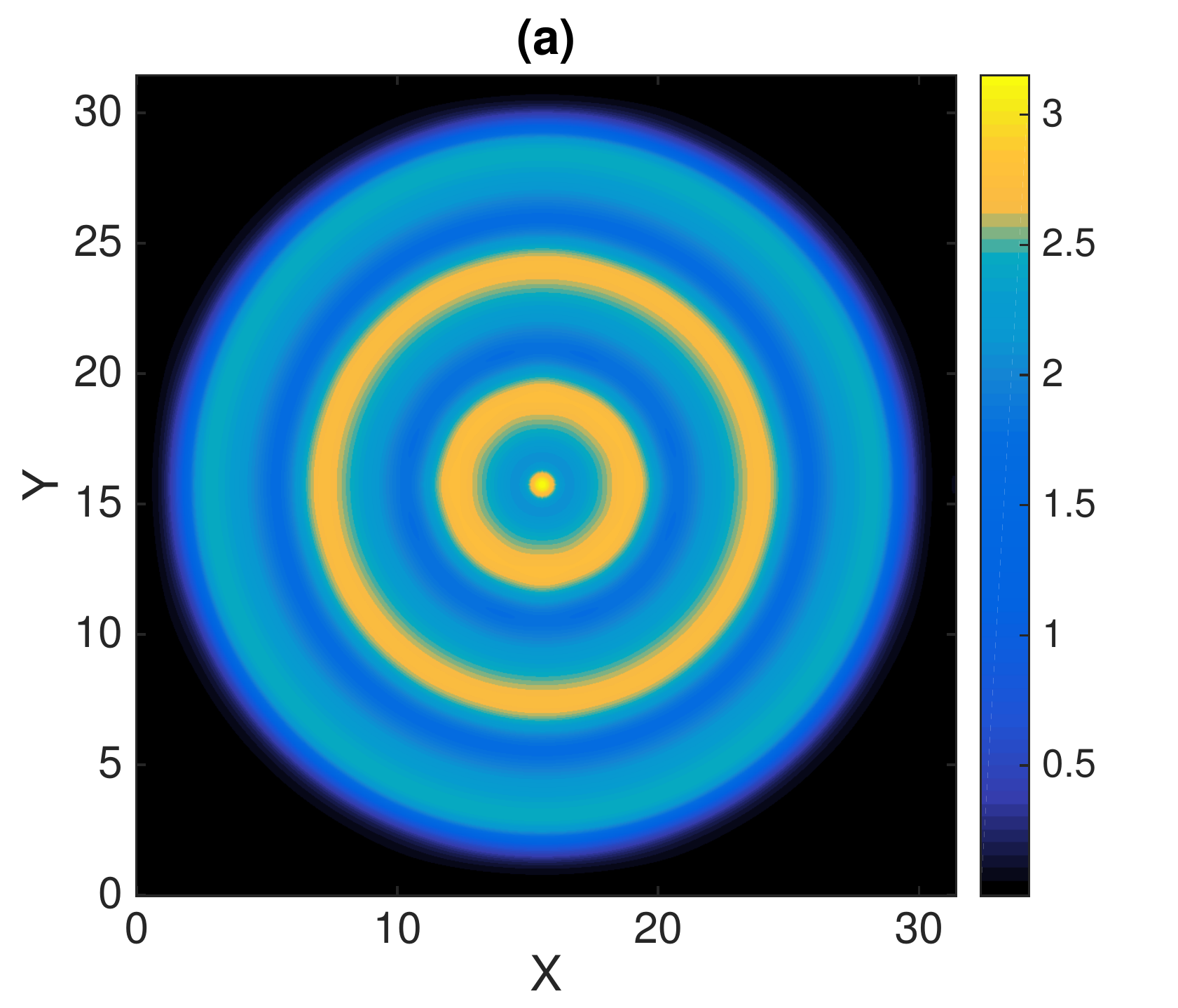}
\includegraphics[width=0.5\linewidth]{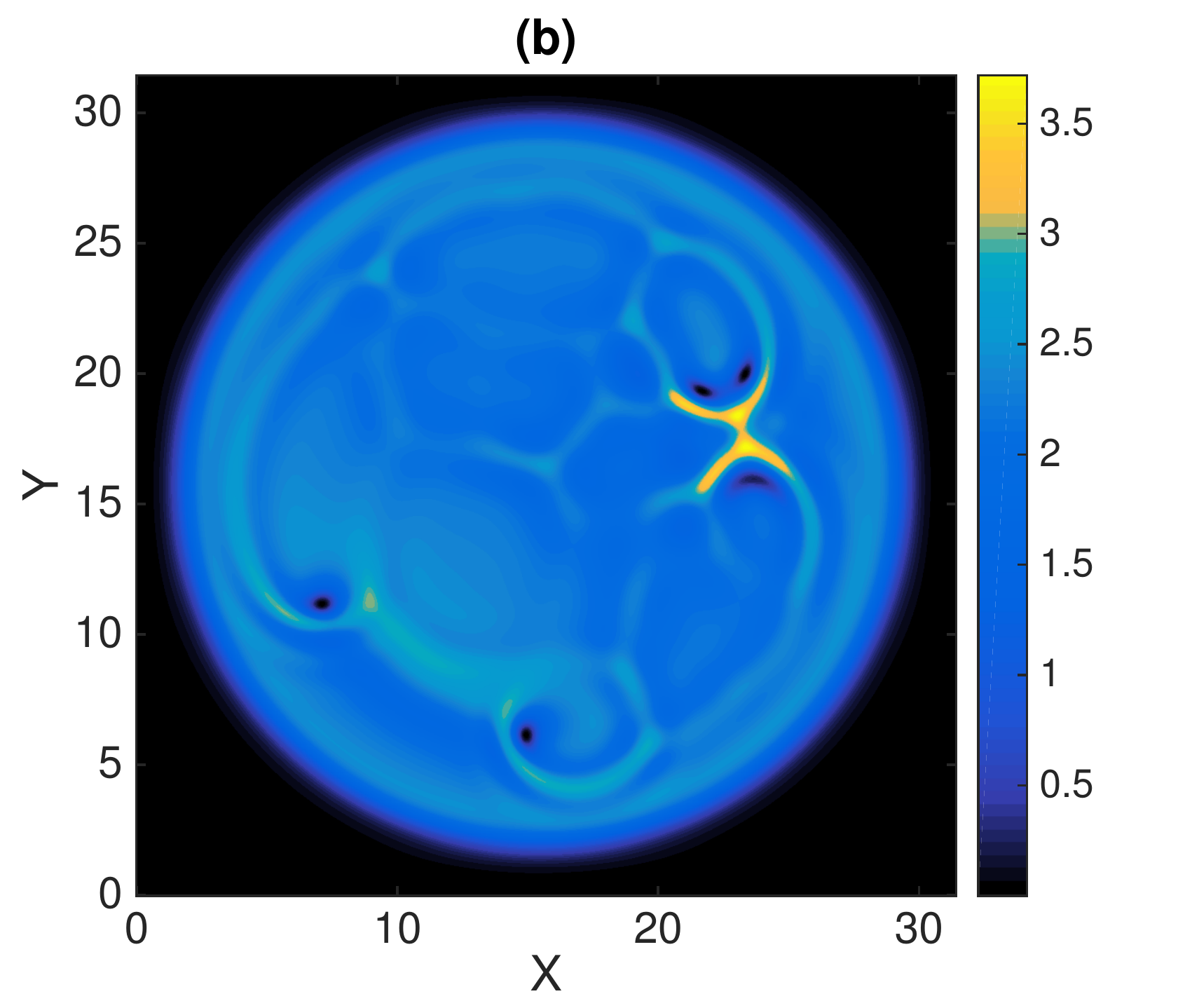}
\includegraphics[width=0.5\linewidth]{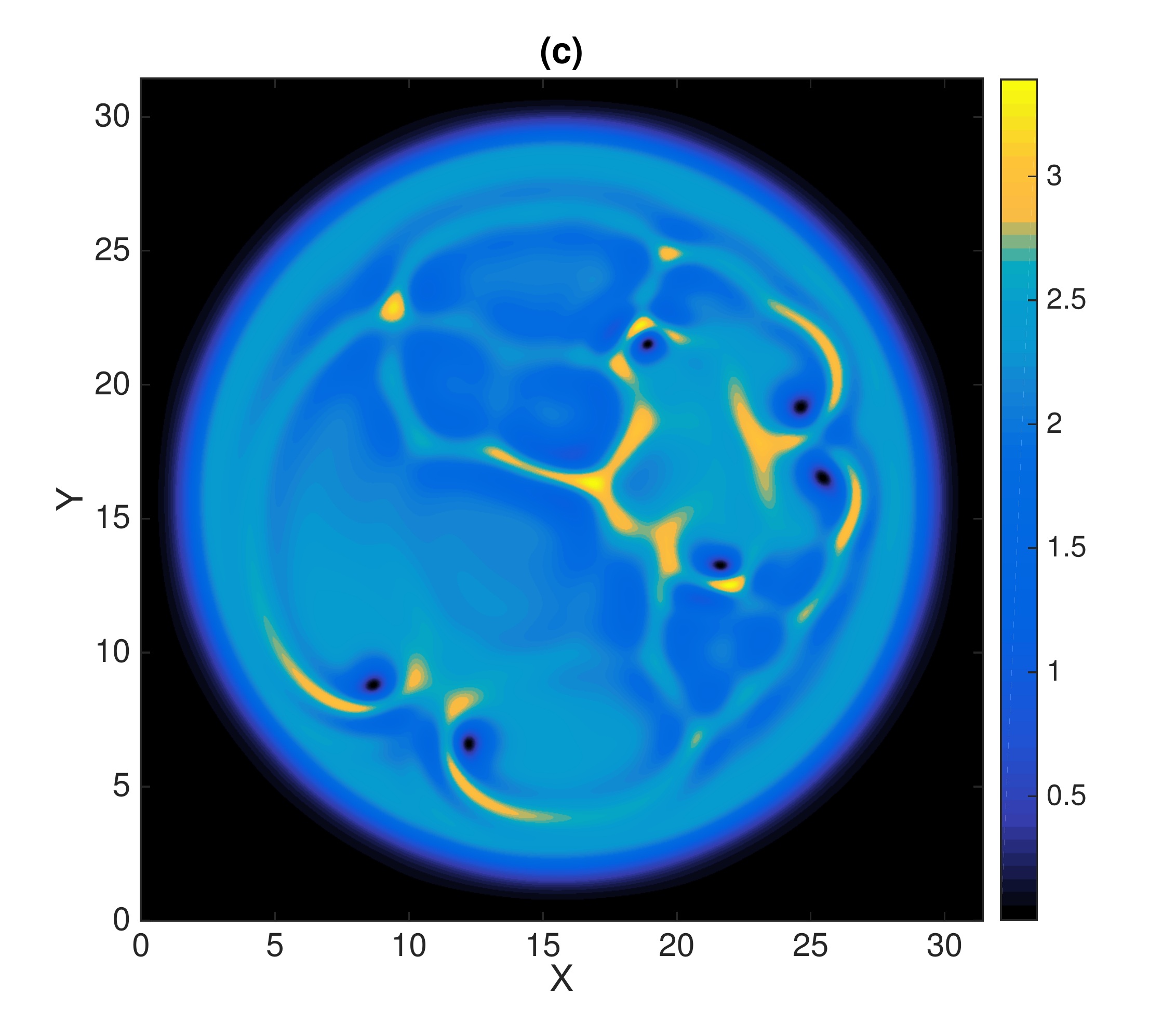}
\includegraphics[width=0.5\linewidth]{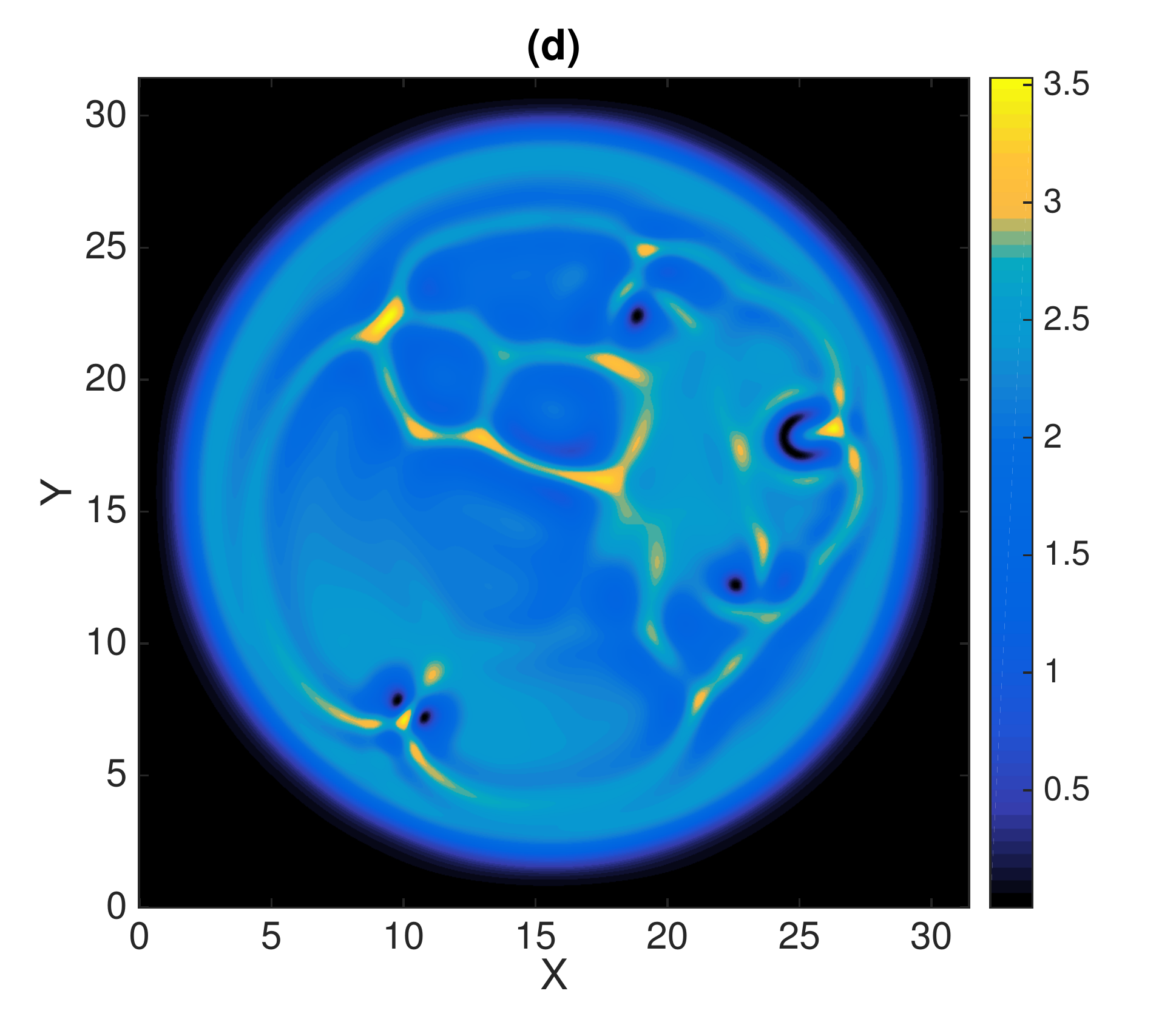}
\caption{Transverse intensity profiles for Eq.~(\ref{laser}) with $\omega=0.53$, $E_{IN}=0.6$ and $P=4$. (a) Periodically modulated steady ring structure. (b)-(d) Spatio-temporal snapshots of optical vortex-mediated turbulence.}
\label{fig3}
\end{figure}

From Fig.~\ref{fig2} (c)-(d) it is clear that in our regime of optical turbulence intensity peaks appear in the middle of regions where the intensity is very low or even zero. In order to clarify the nature of this optical turbulence, we report in Fig.~\ref{fig3} a turbulent regime obtained through the numerical integration of Eq.~(\ref{laser}) and for smaller values of the detuning $\omega$ and injection $E_{IN}$ with respect to the results displayed in Fig.~\ref{fig2}. In this case ($\omega=0.53$, $E_{IN}=0.6$ and $P=4$) there is bistability between the steady spatially modulated ring structure of Fig.~\ref{fig3} (a) and the turbulent regime. By locally perturbing the ring structure, turbulence sets in as shown in Fig.~\ref{fig3} (b)-(d) and remains asymptotically. In this regime, we can clearly identify points of zero intensity where lines of zero real and zero imaginary parts of the field $E$ cross each other. These zero-intensity points are optical vortices \cite{Coullet89} and we observe that they appear in couples of opposite topological charge (see the four vortices on the right of Fig.~\ref{fig3} (b), the first couple being at the top and the second at the bottom) and they also disappear in collisions of oppositely charged pairs (see events in Fig.~\ref{fig3} (d) on both right and left).

The turbulent states that we observe in Figs.~\ref{fig2} and \ref{fig3} appear to be examples of defect mediated turbulence (DMT), a configuration first described by Coullet, Gil and Lega in 1989 \cite{Coullet89b} in the complex Ginzburg-Landau (CGL) equation
\begin{eqnarray}
\label{cgl}
\frac{\partial E}{\partial t} &=& (P-1+i\omega)\;E - \left(1+i\alpha \right) |E|^2 \; E + \left(1+i\beta \right) \nabla^2 E
\end{eqnarray}
where the control parameters are the pump $P$ and the detuning $\omega$ as well as $\alpha$ and $\beta$ that regulate the ratio between dispersive and absorptive nonlinearity and between diffraction and diffusion, respectively. In particular, Coullet, Gil and Lega describe the onset of DMT when the condition $(1+\alpha\beta)< 0$ is met and the defects are pointwise vortices. %
The CGL equation looks quite similar to our Eq. (\ref{laser}) in the absence of injection and of Swift-Hohenberg terms. 
However, since it is not possible to properly define the coefficients $\alpha$ and $\beta$ in Eq.(\ref{laser}), the condition for 
the occurrence of DMT cannot be met in Eq.(\ref{laser}) in the absence of injection. This is probably why DMT has never 
been observed experimentally in nonlinear optics and, in particular, in lasers without injection.

One interesting question that we would like to address in this article is: does the external forcing due to the optical injection help or hinder DMT? In the literature, there are contradictory statements. On one side it is clear that stationary optical vortices are prevented from forming by the presence of the external forcing $E_{IN}$ as it breaks the phase invariance $\exp(i\phi)$ of the CGL equation (\ref{cgl})  \cite{Coullet92}. On the other side, Chat\'e, Pikovsky and Rudzick \cite{Chate99} demonstrated that DMT survives when perturbing the CGL equation with the term $E_{IN}$ in a regime where the condition $(1+\alpha\beta)< 0$ is satisfied. The cases that we have presented in Figs.~\ref{fig2} and \ref{fig3}, however, do not satisfy the condition of Ref. \cite{Chate99} because our unperturbed laser equations never satisfy the phase-to-amplitude instability condition $(1+\alpha\beta)< 0$. The original question above therefore remains unanswered for the laser and SROPO cases. To answer it we need to investigate in more detail the instabilities that lead to the appearance of the turbulent regime displayed in Fig.\ref{fig2} (c)-(d) and Fig.\ref{fig3} (b)-(d).

As shown in \cite{Gibson16}, spatially modulated stationary structures such as those displayed in Figs.~\ref{fig2}~(b) and \ref{fig3}~(a) correspond to phase bound solutions since the variations of the phase of the field $E$ when moving across the spatial variable are limited to a finite arc. In the case of Eq.~(\ref{laser}) the arc is very close to the limit cycle solution of equation \cite{Mayol02,Erneux10}
\begin{eqnarray}
E_{LL}(t) &=& A_0 \left[ \cos(\Omega t+\pi) + i  \sin(\Omega t+\pi) \right]\, , \nonumber \\
\Omega &=& \sqrt{\omega^2-\omega_L^2} \;\; ;\;\;\;\;   \omega_L=E_{IN}/A_0 \;\;;\;\;\;\;
A_0=\sqrt{\frac{3(P-1)}{P}} \; .
\label{limitcycle}
\end{eqnarray}
Note that when $\Omega$ is real, i.e. in the absence of locked states, the trajectory (\ref{limitcycle})
is the phase-drift solution of the Adler equation $d_t \phi = \omega -\omega_L \sin(\phi(t))$ \cite{Adler46}. When the spatially modulated stationary states become unstable, we first observe a phase instability where the new field amplitudes remain close to the limit cycle (\ref{limitcycle}) but the phase spreads around the full circle. After this phase spreading, we observe an amplitude instability with large fluctuations away from the limit cycle $E_{LL}(t)$ (\ref{limitcycle}). These fluctuations progressively approach points of zero-intensity where both the real and imaginary parts of the field $E$ are simultaneously zero. It is exactly at these points that we observe the nucleation of pairs of vortex defects (see for example Fig.\ref{fig3} (b)). These vortices then move around the transverse space and annihilate vortices of opposite topological charge. Once  the amplitude instability driven by $E_{IN}$ takes place, the nucleation and annihilation of optical vortices keeps happening and the DMT regime sets in asymptotically.

We have now established that regimes of DMT are commonplace in our equations (\ref{laser}) and (\ref{sropo}).  This is somewhat surprising since we have external forcing and we cannot meet the condition $(1+\alpha\beta)< 0$ in the unperturbed systems. It is important to note that it is exactly the external forcing that induces first phase and then amplitude instabilities for the generation and stabilisation of DMT. A few important considerations are in order:
1) Our DMT is not related to laser relaxation oscillations because we have considered class A instead of class B lasers. For the effect of relaxation oscillations on rogue waves (RWs) in forced laser systems see Section \ref{sec:0d} and \ref{sec:1d}. Of course, when studying class B lasers with injection in 2D, our DMT may resonate with relaxation oscillations and lead to wild dynamical behaviour.
2) Although DMT can have effects in the dynamics of 1D systems \cite{Chate99}, full interaction, nucleation and annihilation of vortex couples can only be properly described in systems with 2D.
3) This mechanism of vortex turbulence is at the base of the generation of RWs in the 2D externally driven systems described by the spatio-temporal dynamics of Eqs. (\ref{laser}) and (\ref{sropo}), as demonstrated below.

\begin{figure}[htb]
\centering
\includegraphics[width=0.6\linewidth]{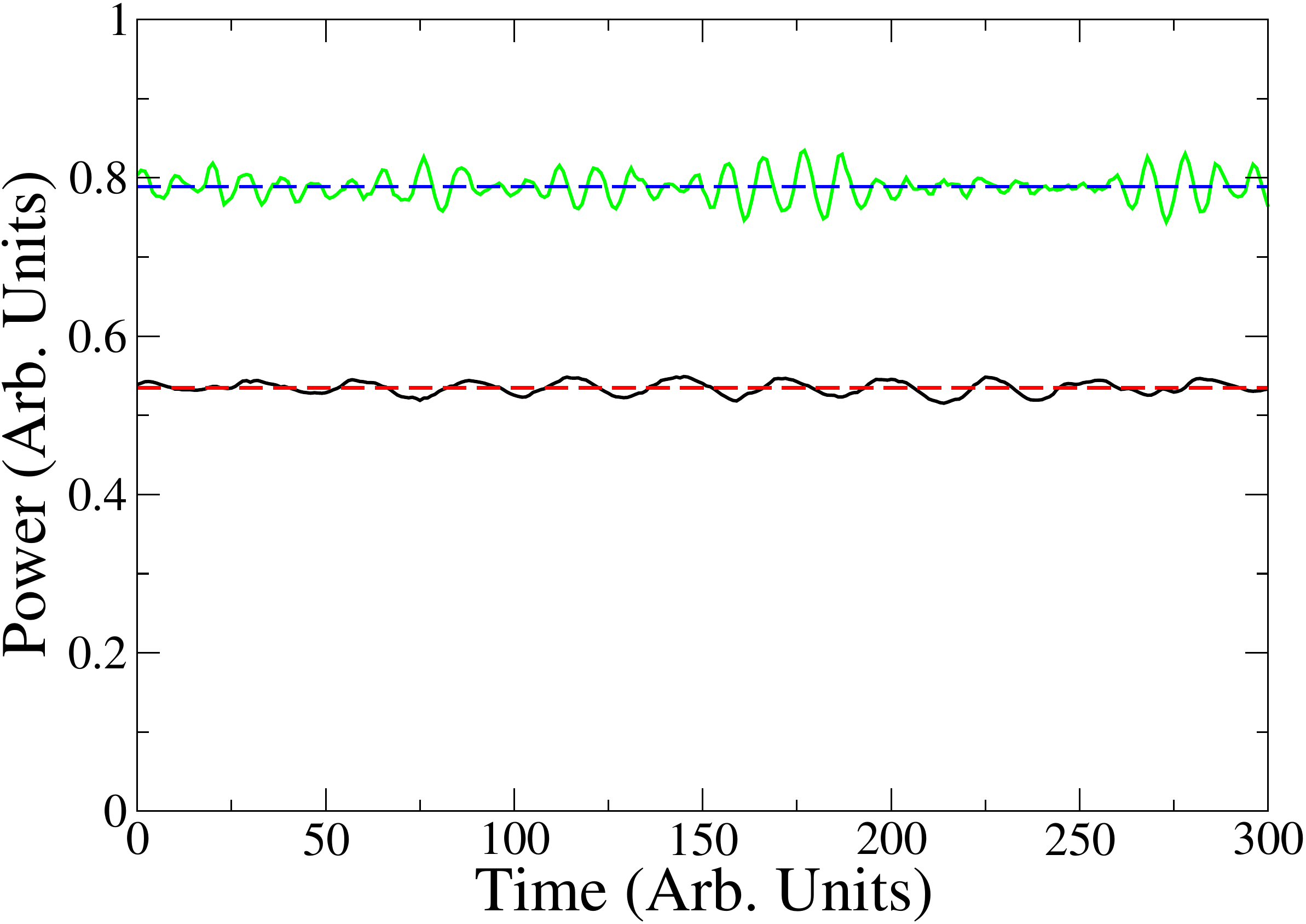}
\caption{Time evolution of the transverse power for Eq. (\ref{laser}) (lower black curve with $\omega=0.53$, $E_{IN}=0.6$, $P=4$ and $\Gamma=0$) and Eq. (\ref{sropo}) (upper green curve with $\omega=1$, $E_{IN}=1.2$, $P=4$ and $\Gamma=0$) during vortex turbulence. The horizontal dashed lines are the corresponding averaged values of the power.}
\label{fig4}
\end{figure}

\begin{figure}[htb]
\centering
\includegraphics[width=0.9\linewidth]{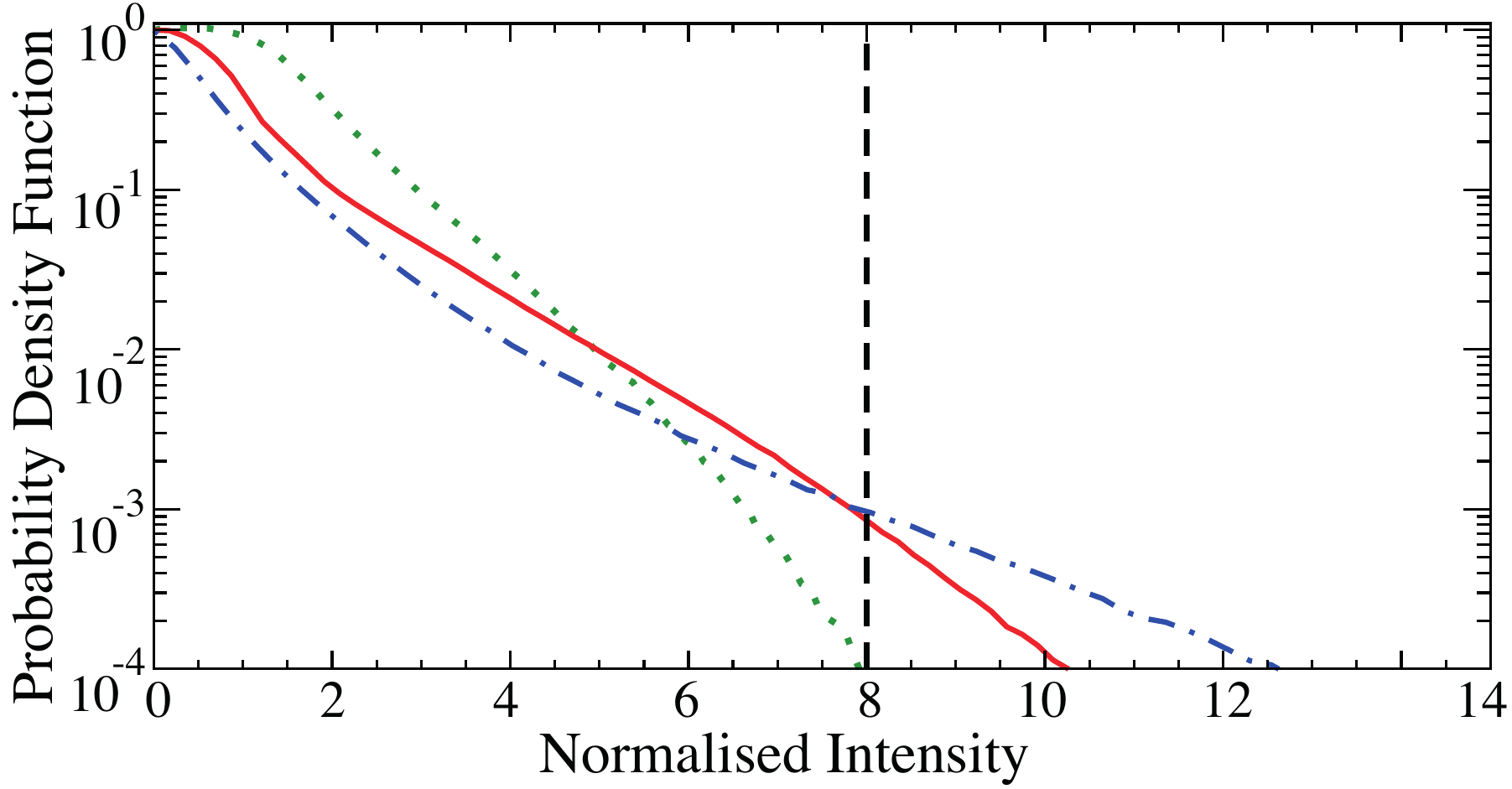}
\caption{Intensity PDFs for Eq. (\ref{laser}) with $P=2, \omega=0.3$, $E_{IN}=0.24$, $\Gamma=0$ (green dotted curve) and with $P=8, \omega=2.4$, $E_{IN}=3.40$, $\Gamma=0$ (red solid curve) and for Eq. (\ref{sropo}) with $P=8,  \omega=1$, $E_{IN}=1.48$, $\Gamma=0$ (dot-dashed blue curve). The vertical black line is the standard threshold for defining waves as an extreme event \cite{Dudley14}. }
\label{fig5}
\end{figure}

Systems  (\ref{laser}) and (\ref{sropo}) are outside thermodynamic equilibrium, do not display relaxation oscillations and present a delicate balance between the energy input and the losses (i.e. between the pump $P$, injection $E_{IN}$, nonlinearity and output mirror losses). For example, during the turbulent evolution, we observe that in spite of large intensity variations locally, the total power displays only small fluctuations around an averaged value as seen in Fig. \ref{fig4}.
In the cases displayed in Fig.~\ref{fig4} the power fluctuations have a standard deviation of only $8$~$10^{-3}$ in the laser case and $1.2$~$10^{-2}$ in the SROPO case. Note also that in the laser case the average power is very close to an analytical estimate of $\pi w_0^2 A_0^2$ where $w_0$ is the beam width of the input pump and $A_0$ is given in (\ref{limitcycle}). The fact that the turbulent state maintains an almost constant power in the presence of moving vortices of zero intensity implies the simultaneous appearance of large amplitude spikes. If the vortex density is large, multi-vortex collisions can occur with the production of large, short-lived spikes in the intensity (see Fig. \ref{fig2} (c) and (d)). Short-lived large intensity spikes are rare but possible events, fitting the characteristics of RWs. The particular shape and symmetry of these spikes is crucially determined by the number and position of the surrounding vortices, a feature that is unique to our particular mechanism of RW formation. Note that without spatial coupling due to diffraction, no RWs can be observed in systems described by Eqs. (\ref{sropo}) and (\ref{laser}). To characterise our spatio-temporal RWs, we use the well-known probability density function (PDF) of the peak intensities \cite{Dudley14} and focus on the shape of the rare event tail. A well defined change in the curvature of the PDF leading to long statistical tails is an essential and clear-cut signature of the presence of RWs \cite{Dudley14}.
In Fig.~\ref{fig5} we display the progressive change in shape of the PDF when moving to higher values of pump $P$, injection $E_{IN}$ and detuning $\omega$. These changes are due to an increased density of vortices in the regimes of DMT.
At lower values of energy input (lower values of $P$ and $E_{IN}$) (see the green dotted curve in Fig. \ref{fig5}) the wave statistics is well-approximated by a Gaussian fit and the tail is not suitable to describe rare events with extremely large peak intensities.  At higher energy inputs  (see the red solid and the dot-dashed blue curves in Fig. \ref{fig5}) a clear change in the shape of the PDF becomes apparent with the appearance of long non-Gaussian tails that correspond to generation of RWs. In the case of the dot-dashed blue curve in Fig. \ref{fig5} the statistics is very well approximated by a Weibull distribution \cite{Onorato13} given by:
\begin{equation}
f(x) = A \left( \frac{x}{b} \right) ^{c-1} \exp \left( -\frac{x}{b} \right)^c
\label{weibull}
\end{equation}
as shown in Ref.\cite{Gibson16}.

Note that non-Gaussian PDFs cannot be replicated by superpositions of random waves typical of optical speckles \cite{Shvartsman94}. It has to be stressed that in regimes of DMT, the turbulent dynamics of vortices is deterministically driven by the nonlinearity. We also note that RWs in the vortex turbulence demonstrated in Fig. \ref{fig5} are different from those due to vorticity in models of inviscid fluids \cite{Abrashkin13}. Eqs. (\ref{laser}) and (\ref{sropo}) have been shown to be equivalent to the flow of a compressible and viscous fluid with density $\rho=|E|^2$ and velocity {\bf v}=$\nabla \phi$ where $\phi$ is the phase of the field \cite{Brambilla91}. In the case of our forced systems, $\nabla \times${\bf v} remains extremely close to zero in the locations where RWs are observed. We conclude that our RWs are due to the interaction of free vortices in the absence of vorticity.

In conclusion, we have shown a mechanism for producing RWs in the 2D transverse area of externally driven nonlinear optical devices via vortex turbulence. Given the universality of our models, this mechanism should be observable in a large variety of systems. Models of lasers with injected signal, where the invariance of the Adler limit cycle is well known \cite{Oppo86,Mayol02}, can easily be extended to semiconductor media and to class B lasers, thus including the largest majority of solid state lasers. Outside optics, vortex-mediated turbulence without driving has been observed in nematic liquid crystals \cite{Frisch95}, chemical reactions \cite{Ouyang96} and fluid dynamics \cite{Bodenschatz00}. In the unlocked regime of these systems with driving, vortex turbulence can excite RWs and lead to the formation of highly inhomogeneous fields with non-Gaussian statistics.

\section{Conclusion}
In this chapter we have reviewed recent results on the statistical properties of forced oscillatory media with different numbers of spatial dimensions. Although the theoretical models differ in their exact form, they are essentially equivalent and an almost direct comparison can be performed, where the impact of spatial dimensionality can be at least partly addressed. In all three cases heavy tailed statistics have been observed in some specific parametric conditions and dynamical regimes. In the 0-dimensional case, noise can be a cause for rogue waves but purely deterministic dynamics can also be sufficient. The impact of noise has not been analyzed yet in the 1-D and 2-D cases but, as in the 0-D case, rogue waves can be observed as a fully deterministic phenomenon. In the 0-dimensional case the vicinity of an external crisis corresponding to the collision of a chaotic attractor with the stable manifold of the saddle-focus point corresponding to the laser ``low intensity'' state can be at the origin of rogue events.  The corresponding trajectory has also been observed experimentally, and some predictions can be realized due to the strongly dominant role of determinism in the dynamics. In the 1-D system two types of rogue events have been detected in real time and analysed in a spatio-temporal diagram. The first type occurs in a disordered regime where the phase of the laser relative to the forcing evolves erratically but in a mostly bounded way except in some rare occurrences when the electric field completes a full 2$\pi$ rotation around the locked point, implying the existence of a defect. The experimental observations clearly show a focussing phenomenon taking place over several round-trips during the formation of a rogue event which can be interpreted as an unstable phase soliton, a feature which is obviously absent in the 0-D case. Although a crisis of attractors would be very difficult to identify in this partial differential equations model, it is notable that the emergence of extreme events is related to the system approaching the saddle point corresponding to the laser ``low-intensity'' state, suggesting that this point plays a similar role to the one it plays in the 0-D case. The second type of extreme events on the contrary is essentially related to the presence of spatial degrees of freedom since large statistical deviations occur when (otherwise stable) localized wave packets partially destructively collide with transient pulses. In the 2-D case rogue waves are found in a regime of defect mediated turbulence. In this case, destructive collisions between vortices of opposite charge are at the origin of the extreme events and they can not be found in 0-D or 1-D systems.

\bigskip

\textbf{Acknowledgments}. CM was supported in part by the Spanish Ministerio de Ciencia e Innovaci\'on through project FIS2015-66503-C3-2-P (MINECO/FEDER) and the ICREA ACADEMIA program, Generalitat de Catalunya, Spain.

\end{document}